\documentclass[12pt,preprint]{aastex}

\renewcommand{\vec}[1]{\mathbf{#1}}
\newcommand{\eqref}[1]{(\ref{#1})}
\newcommand{\grad}{\nabla}
\renewcommand{\div}{\mathbf{\nabla} \cdot}
\newcommand{\rot}{\mathbf{\nabla} \times}
\newcommand{\pp}[2]{\frac{\partial #1}{\partial #2}}
\newcommand{\Alfven}{Alfv\'{e}n }
\newcommand{\JS}{J\"{u}tter-Synge }
\newcommand{\thh}{\tanh (z/\lambda)}
\newcommand{\cs}{\cosh^{-2} (z/\lambda)}

\shorttitle{Relativistic Reconnection \& Drift Kink Instability}
\shortauthors{Zenitani \& Hoshino}

\begin{document}

\title{Particle acceleration and magnetic dissipation
in relativistic current sheet of pair plasmas}

\author{S. Zenitani}
\affil{
NASA Goddard Space Flight Center, Greenbelt, MD 20771;
zenitani@lssp-mail.gsfc.nasa.gov
}
\author{M. Hoshino}
\affil{
Department of Earth and Planetary Science, University of Tokyo,
7-3-1, Hongo, Bunkyo, Tokyo, 113-0033 Japan
}

\begin{abstract}
We study linear and nonlinear development of
relativistic and ultrarelativistic current sheets
of pair ($e^{\pm}$) plasmas
with antiparallel magnetic fields.
Two types of two-dimensional problems are investigated
by particle-in-cell simulations.
First, we present the development of
relativistic magnetic reconnection,
whose outflow speed is on the of the light speed $c$.
It is demonstrated that
particles are strongly accelerated
in and around the reconnection region
and that
most of the magnetic energy is converted into 
a ``nonthermal'' part of plasma kinetic energy.
Second, we present another two-dimensional problem of
a current sheet in a cross-field plane.
In this case, the relativistic drift kink instability (RDKI) occurs.
Particle acceleration also takes place, but the
RDKI quickly dissipates the magnetic energy into plasma heat.
We discuss the mechanism of particle acceleration and
the theory of the RDKI in detail.
It is important that
properties of these two processes are similar
in the relativistic regime of $T\gtrsim mc^2$,
as long as we consider the kinetics.
Comparison of the two processes indicates
that magnetic dissipation by the RDKI
is a more favorable process in the relativistic current sheet.
Therefore, the striped pulsar wind scenario
should be reconsidered by the RDKI.
\end{abstract}

\keywords{acceleration of particles --- magnetic fields --- plasmas --- relativity --- stars: winds, outflows --- pulsars: individual (Crab Pulsar)}

\singlespace

\section{Introduction}

Magnetic reconnection is
an important large-scale processes in space plasmas.
By rearranging the magnetic field topology,
it causes heating and particle acceleration of plasmas
as well as dissipation of the magnetic fields.
Although it has been extensively studied
in wide range of solar terrestrial sites---
stellar and solar flares \citep{parker},
the planetary magnetospheres \citep{dungey,jup98}
and
solar wind \citep{phan06}---
it has also been discussed
in high-energy astrophysical contexts such as
the magnetized loop of the Galactic center \citep{hey88},
the jets from active galactic nuclei \citep{dimatteo,schopper,lb98,larra03},
quite probably gamma-ray bursts \citep{drs02}, and
pulsar winds \citep{michel82,coro90,michel94}.
In these situations, reconnection often takes place
in relativistic electron-positron plasmas.
Especially in the Crab pulsar wind,
in the relativistic outflow of pair plasmas from the central neutron star,
relativistic reconnection or relevant processes
have been considered as possible processes to explain
the long-standing ``$\sigma$-problem''
($\sigma$ is the ratio of the Poynting flux energy to the particle kinetic flux);
the plasma outflow is originally Poynting-dominated ($\sigma \sim 10^4$)
close to the neutron star \citep{arons79},
but it should be kinetic-dominated near the downstream termination shock
($\sigma \sim 10^{-3}$; \cite{kc84a}). 
The possible dissipation mechanisms are
instabilities in the striped current sheets;
due to the fast rotation of the central neutron star,
whose magnetic dipole is in the oblique direction,
alternating magnetic fields are strongly ``striped'' near the equatorial plane.
We assume that there are current sheets
between such opposite magnetic field lines,
and magnetic reconnection occurs and dissipates the magnetic energy
there \citep{coro90}.
Based on this idea, magnetic dissipation in the pulsar wind
has been estimated by using a one-dimensional MHD model \citep{lyu01,kirk03},
but they were not so successful.
The main problem is that
basic properties of relativistic reconnection or current sheet processes
are still unclear.

Until recent years, there have been few
theoretical studies on relativistic reconnection.
\citet{zelenyi79} studied the relativistic kinetic description of
the tearing instability,
which is the most important instability in the reconnection context.
\citet{bf94b} studied the steady state reconnection models
in relativistic pair plasmas. They argued that
the reconnection rate becomes closer to the unity due to Lorentz effects,
and therefore, faster energy conversion is possible.
These steady models are further investigated by several authors \citep{lyutikov03,lyu05},
although there is controversy about
whether inflow speed can be relativistic or not.
On the viewpoint of particle acceleration,
\citet{rom92} studied the particle orbits in the reconnection field model,
and they obtained an energetic energy spectrum of pair plasma,
which is approximated by the power-law distribution with an index of ${-1.5}$.
\citet{larra03} obtained the power-law index of $-1$,
assuming that particle motion is restricted in the neutral plane.
In a fully self-consistent way, relativistic magnetic reconnection
has recently been explored by particle-in-cell (PIC) simulations
\citep{zeni01,zeni05b,claus04}.
\citet{zeni01} demonstrated that 
an enhanced $dc$ acceleration takes place in and around the $X$-type region,
due to the reconnection electric field $E_y$.
This acceleration is so strong that
a power law index of the energy spectrum around the acceleration site is
on the order of $-1$ \citep{zeni01},
and the energy spectrum over the whole simulation domain
is approximated by the power-law with an index of $-3$ \citep{claus04}.
Using relativistic resistive MHD code,
\citet{naoyuki06} presented Petschek reconnection in the mildly relativistic case.

The current sheet configuration
with antiparallel magnetic field lines
is also influenced by cross field instabilities,
whose wavevectors are perpendicular to the reconnection plane. 
In geophysical contexts, 
the lower hybrid drift instability (LHDI) \citep{krall71,dav77}
may be of importance, and
it leads to transport of plasma heat \citep{huba81}
and the fast triggering of magnetic reconnection \citep{shino05,tanaka06}.
The Kelvin-Helmholtz instability (KHI) \citep{yoon96,shino01},
which arises from the velocity shear
between the fast-drifting plasmas in the current sheet and the background plasmas,
is of importance because of its significant modulation,
and it enhances the magnetic diffusion rate by the LHDI \citep{shino01}.
Recently the drift kink instability (DKI) \citep{zhu96,prit96}
and its cousin mode of the drift sausage instability (DSI)
\citep{bu99,yoon01,silin03} have been introduced.
The DKI is a long-wavelength electromagnetic mode,
which is driven by the fast-drifting plasmas in a thin current sheet and
it quickly grows when the mass ratio of the positively charged particles
to the electrons is small \citep{dau98,dau99}.
Therefore the DKI smears out in geophysical ion-electron situations
but the nonlinear development of the LHDI
may lead to the current sheet modulation by the DKI \citep{hori99}.
On the contrary, the DKI quickly grows
in relativistic pair plasmas in which the mass ratio is the unity,
while the LHDI smears out.
\citet{zeni03,zeni05a} investigated
the DKI in relativistic current sheets of pair plasmas,
and they found a $dc$ particle acceleration along the neutral sheet
in the nonlinear stage of the relativistic drift kink instability (RDKI).
Because of its fast growth rate, the RDKI is
one of the most likely processes in relativistic current sheets.

The purpose of this paper is
to investigate basic physical properties of
reconnection or reconnection-related processes
in relativistic pair plasmas.
Two types of two-dimensional problems,
relativistic magnetic reconnection and the RDKI,
are demonstrated
by using fully electromagnetic PIC simulations.
In addition, based on the obtained theories
we discuss possible application to the Crab pulsar wind problem.

This paper is organized as follows.
In \S 2 we describe our simulation setup.
In \S 3 a two-dimensional simulation study of
relativistic magnetic reconnection is presented.
In \S 4 another two-dimensional study of the RDKI is presented.
In \S 5 we compare the two results and
discuss the possible application to the pulsar winds.

\section{Simulation}

\subsection{Simulation method}

The simulation is carried out by
a three-dimentional electro\-magnetic PIC code.
In this code we solve the following basic equations:
the relativistic equation of motion
\begin{eqnarray}
\frac{d}{dt}
( m_j \gamma_j \vec{v}_j )
&=&
q_j \Big( \vec{E} +  \frac{\vec{v}_j}{c} \times \vec{B}  \Big)
\\
\frac{d}{dt}
\vec{x}_j
&=&
\vec{v}_j
\end{eqnarray}
and the Maxwell equations
\begin{eqnarray}
\rot \vec{B}
&=&
\frac{4\pi}{c} \vec{j} + \frac{1}{c} \pp{}{t} \vec{E} \\
\rot \vec{E}
&=&
- \frac{1}{c} \pp{}{t} \vec{B} \\
\div \vec{B} &=& 0 \\
\div \vec{E} &=& 4\pi \rho
\end{eqnarray}
where $\vec{x}_j, \vec{v}_j, \gamma_j$, $m_j$, and $q_j$ are
the position, the velocity, the Lorentz factor, the rest mass, and
the charge for the $j$-th super-particle in the simulation system.
The charge density $\rho$ and the current density $\vec{j}$
are obtained by
\begin{eqnarray}
\rho &=& \sum_{j} q_j  S(\vec{x}_j)\\
\vec{j} &=& \sum_{j} q_j \vec{v}_j S(\vec{x}_j)
\end{eqnarray}
where $\sum$ denotes integration in the cell and
$S(x)$ is the shape function of the super-particles.
For simplicity, we do not consider any collisions, radiation,
pair production, or pair annihilation of electron-positron plasmas.

\subsection{Initial configuration}

As an initial configuration,
we use a relativistic Harris configuration \citep{kirk03,Harris}.
In the Cartesian coordinate system,
magnetic field and plasma distribution functions are
\begin{eqnarray}
\vec{B} &=& B_0 \tanh(z/\lambda) \hat{\vec{x}},\\
f_{s} &=& \frac{n_0\cosh^{-2}(z/\lambda)}{4\pi m^2cT K_2(mc^2/T)}
\exp\big[
\frac{ -\gamma_s (\varepsilon - \beta_s mc u_y) }{ T }
\big] \nonumber \\
&+& \frac{n_{bg}}{4\pi m^2cT_{bg} K_2(mc^2/T_{bg})} \exp\big[-\frac{\varepsilon}{T_{bg}} \big],
\end{eqnarray}
where $\lambda$ is the thickness of the current sheet,
the subscript $s$ denotes the species
(plus signs for positrons and minus signs for electrons),
$n_0$ is the plasma number density of the current sheet
in the proper frame,
$T$ is the plasma temperature including the Boltzmann constant,
$K_2(x)$ is the modified Bessel function of the second kind,
$c\beta_s$ is the drift speed of the species, and $u$ is the four-velocity.
Throughout this paper we set $\beta_+=0.3$ and $\beta_-=-0.3$ 
and therefore $\gamma_{\beta}=1.048$.
The $n_{bg}$ is the number density of background plasmas, and
$T_{bg}$ is its temperature including the Boltzmann constant. 
The pressure balance condition and the current condition
are satisfied in the equilibrium state;
\begin{equation}\label{eq:pb}
B_0^2 / 8\pi = 2d_0 T/\gamma_{\beta} = 2n_0 T
\end{equation}
\begin{equation}\label{eq:curr}
c B_0 / ( 4\pi \lambda ) = \sum_s \gamma_s q_s n_s v_s = 2en_0 \gamma_{\beta} c \beta = 2ed_0 c \beta,
\end{equation}
where $d_0=\gamma_{\beta}n_0$ is the plasma density in the laboratory frame.

In the Harris system, we can take two free parameters.
We employ the temperature $T$ and the drift speed parameter $\beta$. 
The temperature $T$ is a measure of the relativity in this study.
In the nonrelativistic regime,
one can obtain the typical \Alfven speed $V_A$ in the system
from equation \eqref{eq:pb}.
\begin{equation}
V_A \sim \frac{B_0}{\sqrt{4\pi m (2n_0)}} = \sqrt{2}~ \Big(\frac{T}{mc^2}\Big)^{1/2}~ c
\end{equation}
Therefore,
the typical \Alfven speed becomes on the order of $c$ and then
several relativistic effects appear, when $T \gtrsim mc^2$. 
The drift speed parameter $\beta$ also stands for the current sheet thickness.
Combining equations \eqref{eq:pb} and \eqref{eq:curr},
one can obtain the Debye length $\lambda_D$,
\begin{equation}\label{eq:debye}
\lambda_D = \sqrt{\frac{T}{4\pi d_0 e^2 \gamma_{\beta}}} =\beta \cdot \lambda .
\end{equation}
The typical gyroradius $r_L$ in the $T \gg mc^2$ limit can be
approximated as
\begin{equation}\label{eq:rgyro}
r_L \sim \Big(\frac{eB_0}{\gamma mc}\Big)^{-1} c = \frac{(T/\gamma_{\beta})}{mc^2} \frac{mc^2}{eB_0} = (\beta / 2\gamma_{\beta}) \cdot \lambda .
\end{equation}
In the nonrelativistic limit of $T \ll mc^2$,
\begin{equation}\label{eq:gyro}
r_L = \Big(\frac{eB_0}{mc}\Big)^{-1} v_{th} = \frac{mc}{eB_0} \sqrt{\frac{2T}{m \gamma_{\beta}}} 
= \frac{\beta}{\sqrt{2\gamma_{\beta}}} \Big(\frac{T}{mc^2} \Big)^{-1/2} \cdot \lambda ,
\end{equation}
where $v_{th}$ is the thermal velocity of plasmas.

We use the three-dimensional PIC code.
The system size is set to
1024 ($x$) $\times$ 1 ($y$) $\times$ 512 ($z$)
in reconnection studies and
1 ($x$) $\times$ 256 ($y$) $\times$ 512 ($z$)
in drift kink studies.
We consider periodic boundaries in the $x$-, $y$- and $z$-directions.
Since the magnetic field lines change
their directions in a current sheet,
we set two simulation domains in the $z$ direction: 
one domain for the first Harris sheet
and the other for the second Harris sheet,
which has the same physical properties in the opposite directions.
Usually, physical processes are investigated
in the first half (bottom half) of the whole simulation box,
and so the effective size in $z$ is 256 grids.
The typical scale of the current sheet $\lambda$
is set to 10 grids.
Therefore, the boundaries are located
at $x = \pm 51.2 \lambda$ (reconnection) or
at $y = \pm 12.8 \lambda$ (drift kink).
The $z$-boundaries of the main simulation domain are
located at $z = \pm 12.8 \lambda$. 
Simulation time is normalized by
the light transit time $\tau_{c}=\lambda/c$.

A list of simulation runs is presented in Table \ref{table}.
Note that $\omega_{c}=eB_0/mc$ means ``unit gyroradius''.
In the relativistic limit, the typical gyroradius
becomes larger by a factor of $T$ (eq. [\ref{eq:rgyro}]).

\section{Relativistic magnetic reconnection}

\subsection{Simulation result}

First, we present our simulation study of
relativistic magnetic reconnection.
The system size is 1024 ($x$) $\times$ 1 ($y$) $\times$ 512 ($z$),
and the physical size of the main simulation box is
$-51.2 < x/\lambda < 51.2$, $-12.8 < z/\lambda < 12.8$.
The plasma temperature is set to $T/mc^2=1.0$,
so that the system's typical \Alfven speed is $V_A \sim 0.53c$.
The background plasma temperature $T_{bg}$ is set to $0.1 mc^2$.
We call this run run R3 in Table \ref{table}. 
In order to trigger an $X$-type neutral line
around the center of the simulation box,
we add artificial electric fields
in the very early stage of the simulation. 
We assume the triggering electric field $E_{trig}$ and 
its maximum amplitude, typical location, and duration time are
$0.3 B_0 ( V_A / c )$, $(x \pm \Delta x,z \pm \Delta z)=(0 \pm 2, 3 \pm 1)$,
and $10 \lambda /V_A$, respectively. 
During $0<t<\tau_{trig}$, we force $E_y \ge E_{trig}$
so that plasmas enter the $X$ point.
Its duration $10 \lambda /V_A \sim 10-15 \tau_c$ is short enough,
compared with the total timescale of the simulation.
Because of this modification
the system slightly gains energy by 0.2\% in the very early stage,
and the total energy is conserved within 0.4\% error
until the end of the simulation at $t/\tau_c=300.0$.

\begin{figure}[htbp]
\begin{center}
\includegraphics[width={\columnwidth},clip]{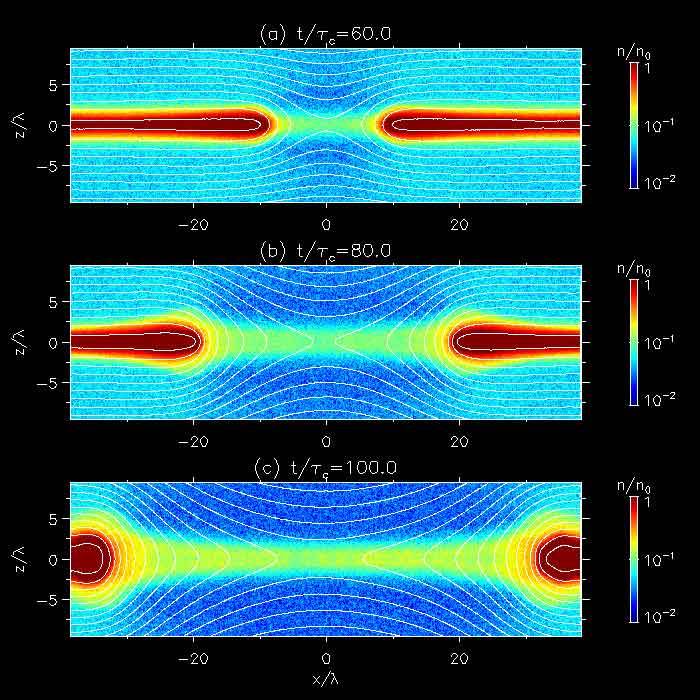}
\caption{
Snapshots of relativistic reconnection at typical stages:
$t/\tau_c=60.0, 80.0$, and $100.0$.
In the left panels, the solid lines represent magnetic field lines
and the color contour shows the density of plasmas
normalized by the original plasma sheet density $\rho_0 = 2\gamma_{\beta}n_0$.
\label{fig:rec_snapshot}}
\end{center}
\end{figure}

\begin{figure}[htbp]
\begin{center}
\includegraphics[width=\columnwidth,clip]{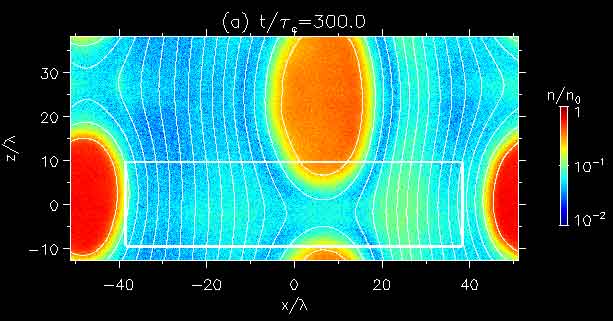}
\caption{
Late-time snapshot at $t/\tau_c=300.0$ of run R3.
Main ($-12.8<z/\lambda<12.8$) and
sub ($12.8<z/\lambda<38.4$) simulation boxes are presented.
The white rectangle shows
the region of interest in Fig. \ref{fig:rec_snapshot}.
\label{fig:rec3}}
\end{center}
\end{figure}

The three panels in Figure \ref{fig:rec_snapshot} present
the snapshots at $t/\tau_c = 60.0, 80.0$, and $100.0$.
The color contour shows the density of plasmas,
which is normalized by the original plasma sheet density $\rho_0 = 2\gamma_{\beta}n_0$. 
The solid lines represent magnetic field lines.
The field lines are originally in the $+x$ direction
on the upside of the current sheet
and the $-x$ direction on the downside of the current sheet.
Because of the triggering field,
the field lines start to reconnect near the center of the simulation box.
Along with the magnetic field lines,
plasmas from the background region come into the $X$-type region 
and then they stream out as reconnection jets 
toward the $\pm x$ directions.
The evolution looks like Sweet-Parker type reconnection
with a flat current sheet structure.
The average velocity of plasmas $\langle\vec{v}\rangle=\int f \vec{v} d\vec{u}/\int f d\vec{u}$
is up to (0.8-0.9)$c$ in the outflow regions
and (0.3-0.4)$c$ in the inflow regions.
The outflow speed exceeds the typical \Alfven speed in the system.
At $t/\tau_c=80.0$ and $100.0$, 
reconnected magnetic field lines ($B_z$) are
swept away from the $X$-type region.
As a result, the field lines are piled up around
$x/\lambda=\pm 20$ or $\pm 30$ in front of the dense plasma regions. 
After $t/\tau_c=100.0$,
the system is influenced by the periodic boundary conditions.
Reconnection outflow jets come closer
to the $x$-periodic boundaries at $x/\lambda=\pm51.2$,
where the plasma density starts to increase.
In addition, the $X$-point starts to sweep
magnetic flux from the other simulation box.
At $t/\tau_c=100.0$, the magnetic field lines at the $X$-point
come from $z/\lambda\sim \pm $12-13 around $x/\lambda = \pm 51.2$.
Note that the $z$-boundaries are located at $z/\lambda=\pm 12.8$.
Dense plasma regions are blown away from the center,
they meet each other near the $x$-periodic boundaries.
Then they merge into a single $O$-point,
and the $O$-point starts to evolve in the vertical ($z$) direction.
After that, the system evolves gradually to
another relaxed state: vertical current sheets.
Figure \ref{fig:rec3} is
a snapshot in the late-time stage at $t/\tau_c=300.0$.
Both main ($-12.8<z/\lambda<12.8$) and
sub ($12.8<z/\lambda<38.4$) simulation boxes are presented.
The white rectangle indicates the region
in the snapshots from Figure \ref{fig:rec_snapshot}.
The magnetic fields are antiparallel;
$B_z<0$ in the left half ($z/\lambda < 0$) and
$B_z>0$ in the right half ($0<z/\lambda$),
but at present we do not observe
the secondary magnetic reconnection
in these vertical current sheets.

\begin{figure}[htbp]
\begin{center}
\includegraphics[width=0.7\columnwidth,clip]{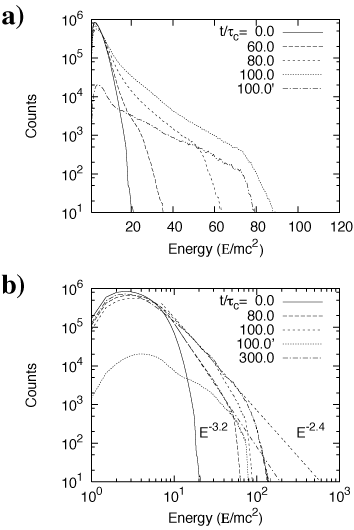}
\caption{\label{fig:recspec}
Particle energy spectra in run R3 are presented
in log-linear and log-log format.
The partial energy spectra at $t/\tau_c=100.0$
is also presented as $t/\tau_c=100.0'$.
}
\end{center}
\end{figure}

The panels in Figure \ref{fig:recspec} are
energy spectra in the main simulation box.
The vertical axis shows the count number of super particles
on a log scale, and the horizontal axis shows particle's energy
which is normalized by the rest-mass energy $mc^2$.
The bottom panel of Figure \ref{fig:recspec}
presents the energy spectra in double-log format.
The spectrum labeled by $t/\tau_c=100.0'$ is
the partial energy spectrum at $t/\tau_c=100.0$,
which is integrated near the central region of $-12.8<x/\lambda<12.8$.
After the reconnection breaks up,
the high-energy tail rapidly grows in time.
This is due to the $dc$ particle acceleration
around the $X$-point \citep{zeni01} and
we discuss the acceleration mechanism in more detail in \S 3.2.
The slope of the spectrum at $t/\tau_c=80.0$
is well approximated by the power-law distribution
with an index of $-3.1$ to $-3.2$
in the range of $10 <\varepsilon/mc^2< 50$.
This is consistent with the other study;
\citet{claus04} obtained a spectral index of $-3$
in their PIC simulations with lower temperature $T/mc^2=0.2$.
At $t/\tau_c=100.0$, the spectrum seems to have a double power-law shape,
whose hardest slope is roughly approximated by the index of $-2.4$.
The partial energy spectrum at $t/\tau_c=100.0$ looks harder.
Finally, the maximum energy is up to $150 mc^2$.
The late-time state at $t/\tau_c=300.0$ also has
its spectral index of $-2.4$.

\begin{figure}[htbp]
\begin{center}
\includegraphics[width={\columnwidth},clip]{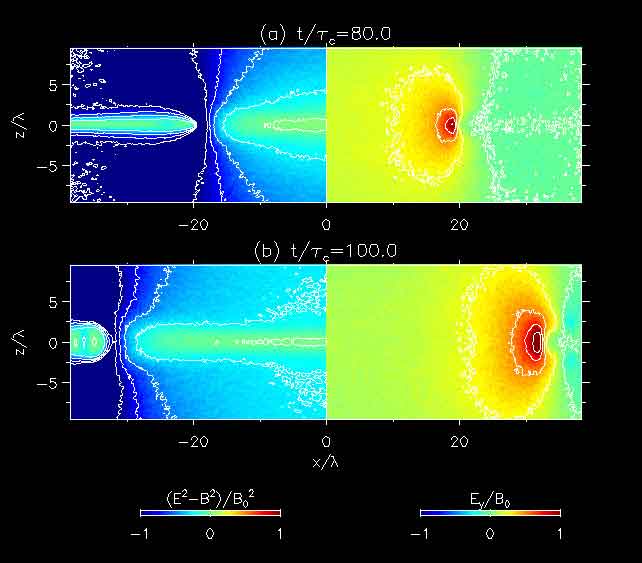}
\caption{
Structure of the electric field in run R3.
The left half displays the Lorentz invariant $E^2-B^2$, normalized by $B^2_0$.
The right half displays the reconnection electric field $E_y$, normalized by $B_0$.
\label{fig:recE}}
\end{center}
\end{figure}

The panels in Figure \ref{fig:recE}
show the electric field structure
at $t/\tau_c=80.0$ and $100.0$.
Because of the symmetry, we present two properties,
$(E^2-B^2)$ in the left halves ($x/\lambda<0$) and 
the reconnection electric field $E_y$ in the right halves ($0<x/\lambda$).
The white contour lines are drawn by $0.25$.
Since the reconnection magnetic fields ($B_x$ or $B_z$) are small
near the $X$-type region
and since the reconnection electric field $E_y$ is finite,
$(E^2-B^2)$ becomes positive inside the narrow sheet region
around the $X$-type region; $-10<x/\lambda<10$ at $t/\tau_c=80.0$ and
$-15<x/\lambda<15$ at $t/\tau_c=100.0$.
The reconnection electric field $E_y$ is
almost constant around the $X$-type region:
$E_y/B_0\sim0.2$ at $t/\tau_c=80.0$
and $E_y/B_0\sim0.15$ at $t/\tau_c=100.0$.
There are strong peaks around $x/\lambda = \pm 20$ at $t/\tau_c=80.0$,
$x/\lambda = \pm 30$ at $t/\tau_c=100.0$.
These regions are identical to the magnetic pile-up regions,
where the reconnected magnetic field lines are compressed
in front of the dense current sheet regions.
The electric fields $E_y$ are enhanced
due to the motional electric field of the magnetic piled-up flux. 
The other two components ($E_x$ and $E_z$) are negligible.

\subsection{Particle acceleration}

\begin{figure}[htbp]
\begin{center}
\includegraphics[width=\columnwidth,clip]{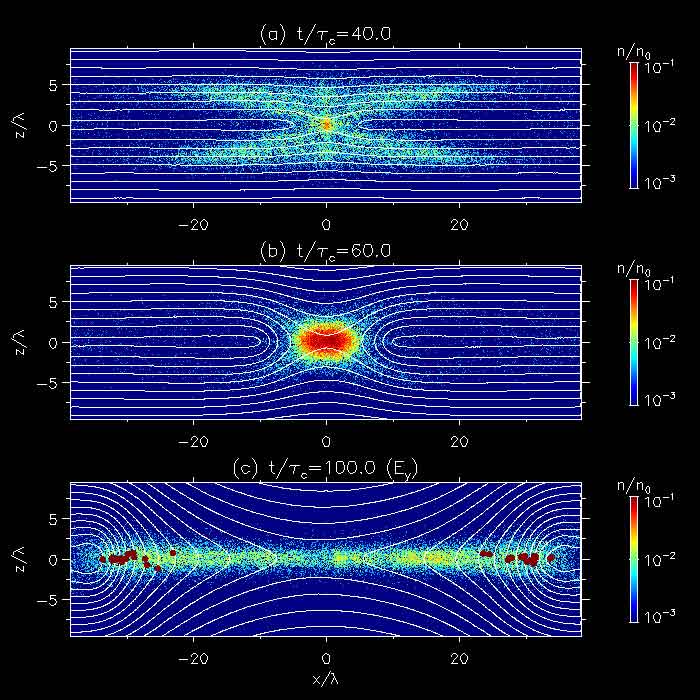}
\caption{
Spatial distributions of accelerated particles
whose energy exceeds $50 mc^2$ at $t/\tau_{c}=100.0$.
White lines show magnetic field lines.
The red marks in the bottom panel show
the spatial distribution of highly accelerated particles,
whose energy exceeds $90 mc^2$.
\label{fig:dist}}
\end{center}
\end{figure}

Next, we look at the particle acceleration
in relativistic reconnection.
In order to analyze how and where
the nonthermal particles are accelerated,
we pick up high-energy particles
whose energy exceeds $50 mc^2$ at $t/\tau_{c}=100.0$.
The panels in Figure \ref{fig:dist} show
their spatial distribution at three different stages.
At $t/\tau_c=40.0$, some particles are near the $X$-type region
and other particles are widely scattered in the background region.
Importantly, they stay around or move into the $X$-type region
at $t/\tau_c=60.0$, when the reconnection breaks up.
The nonthermal tail in the energy spectra starts to grow in this stage.
As the reconnection evolves,
they are spread over the thin current sheet,
and then some of them are ejected from the reconnection region.
The acceleration site is obviously
the central region in and around the $X$-type region.
Looking at the electromagnetic fields
near the acceleration site in panels in Figure \ref{fig:recE},
we note that the electric fields are relatively stronger than the magnetic fields.

We call the central region that satisfies $(E^2-B^2)>0$
the ``acceleration region'' (AR),
where the electric field is dominant.
In two-dimensional reconnection,
$(E^2-B^2)>0$ is equivalent to
our old definition of the AR: $|E_y|>|B^2_x+B^2_z|$ in \citet{zeni01} (hereafter ZH01).
Such an AR does not have a de Hoffman-Teller frame.
For example, let us consider an electromagnetic field $\vec{B}( 0, 0, B_z ), \vec{E}( 0, E_y, 0 )$ near the $X$-points along the neutral sheet ($z=0$).
The fields can be transformed into
\begin{equation}
\vec{B}( 0, 0, B_z ), \vec{E}( 0, E_y, 0 ) \rightarrow
\left\{ \begin{array}{ll}
\vec{B}' ( 0, 0, B_z/\gamma_{v_1} ), \vec{E}' ( 0, 0, 0 ) & (|E_y| < |B_z|) \\
\vec{B}' ( 0, 0, 0 ), \vec{E}' ( 0, E_y / \gamma_{v_2}, 0 ) & (|E_y| > |B_z|) \\
\end{array} \right.
\label{eq:lt}
\end{equation}
where the prime is for the transformed properties,
$v_1/c=E_y/B_z$ and $v_2/c=B_z/E_y$ are the transformation speeds,
and $\gamma_{v_1} and \gamma_{v_2}$ are the relevant Lorentz factors.
Generally, when the Lorentz invariant $(E^2-B^2)$ is positive,
the electric field cannot be removed and
the particle motion in the AR is controlled
by the electric field in the transformed frame.
In the observer's frame,
particles are driven into some $y$-aligned direction,
gaining their energy through the reconnection electric field $E_y$. 
So, once particles enter the acceleration site,
they are driven by the electric field $E'_y$ inside the AR
and
they travel through (relativistic) Speiser orbits \citep{Speiser}
outside the AR.
The AR looks thinner in $z$ compared with ZH01,
but the thickness of the AR does not provide significant change
because the magnetic field $B_x$ confines
particles near the neutral sheet in a reconnection configuration.
In ZH01, we estimate that
the energy distribution in the AR can be approximated
by the power law distribution, whose index is on the order of $-1$.
Although it is difficult to evaluate the partial energy spectra
in the AR by power law, the index is between $-2$ and $-1$
in Figure \ref{fig:recspec}.

\begin{figure}[htbp]
\begin{center}
\includegraphics[width={0.7\columnwidth},clip]{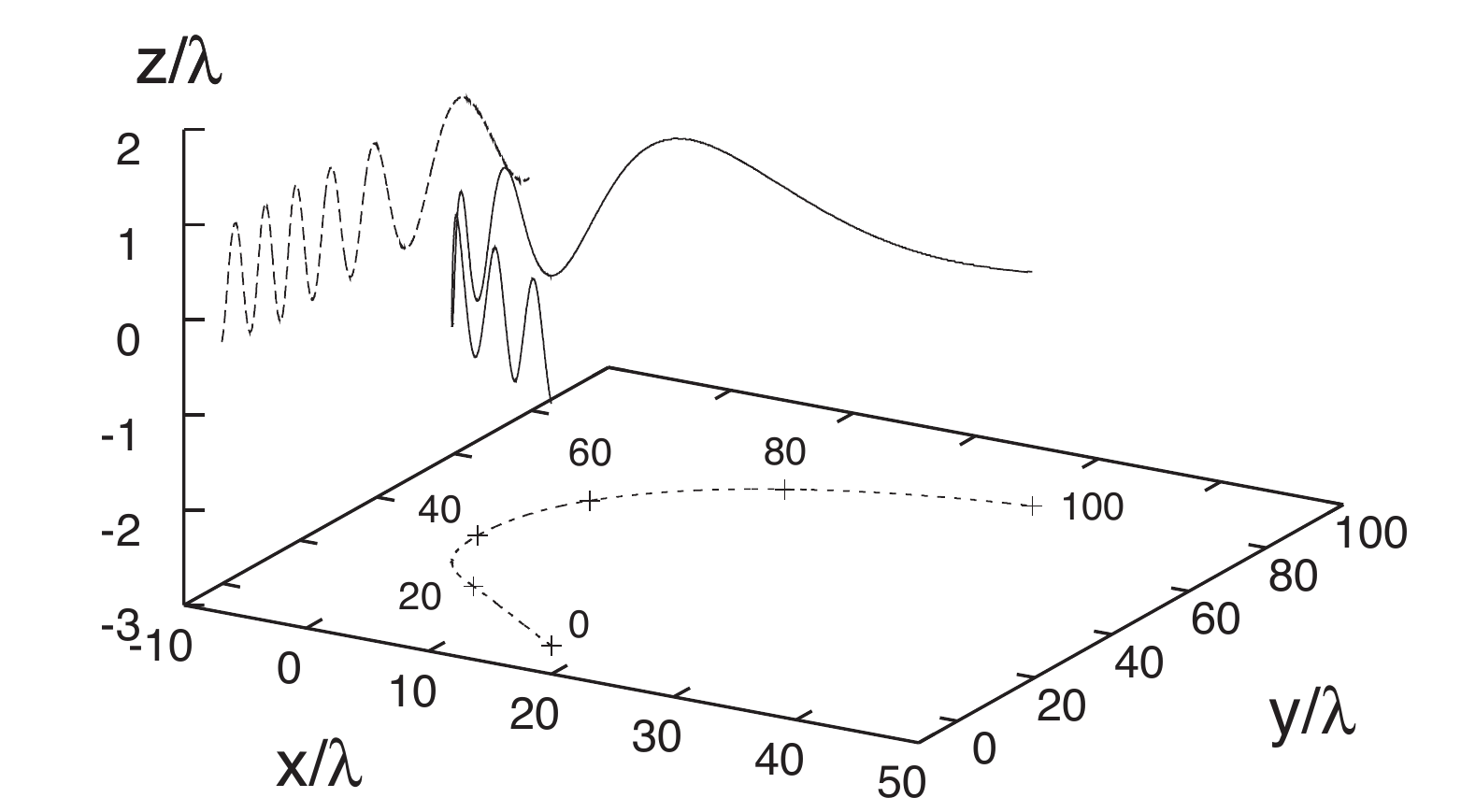}
\caption{Particle trajectory (\textit{solid line}) of a highly accelerated positron, its $xz$- and $xy$-projections (\textit{dashed lines}) are presented.
Labels in the $xy$ plane show the relevant times.
\label{fig:orbit}}
\includegraphics[width={0.8\columnwidth},clip]{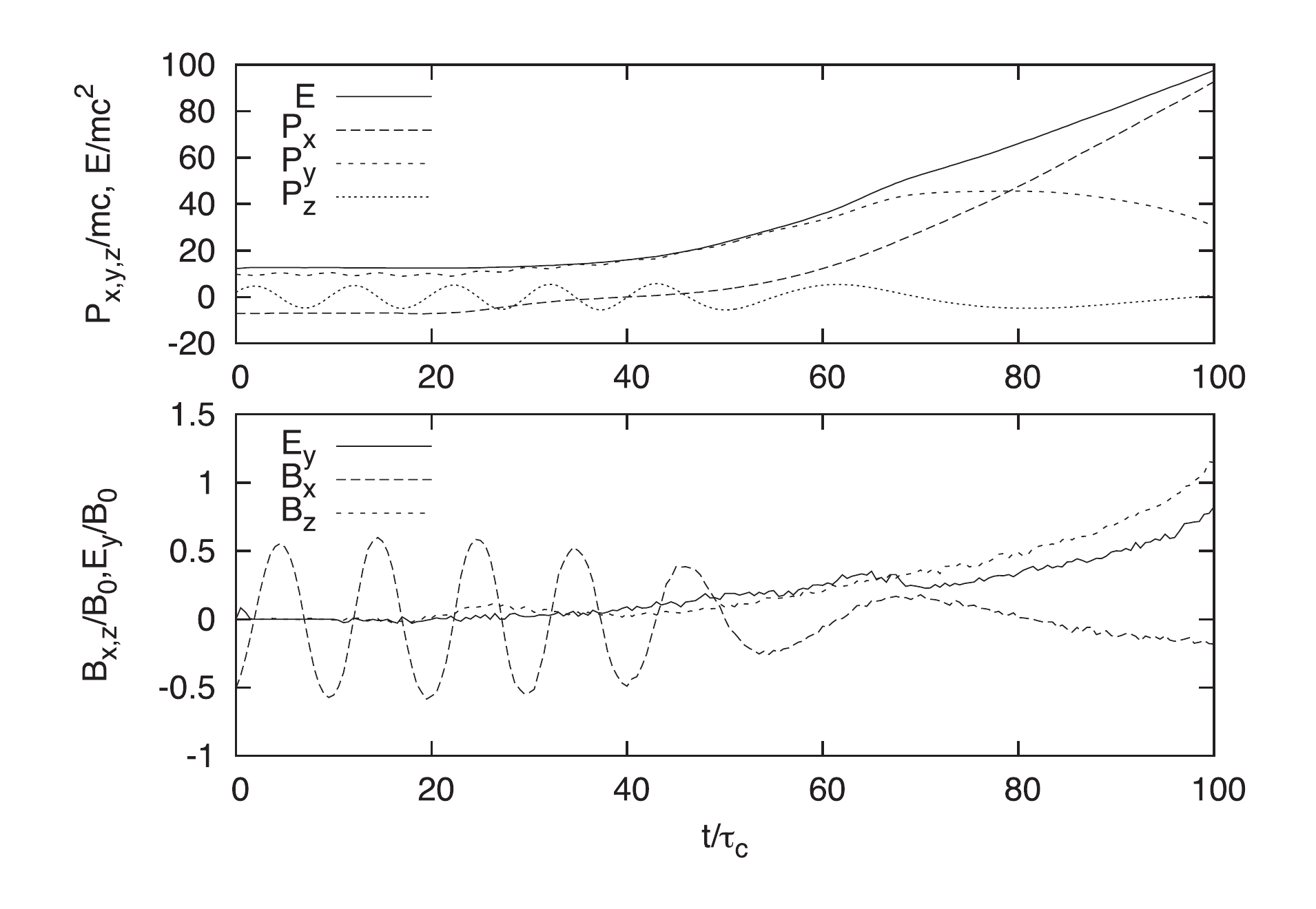}
\caption{
Time history of the particle's energy, momentum and
the electromagnetic field at the positron's position
in the simulation frame.
\label{fig:orbit2}}
\end{center}
\end{figure}

In the bottom panel of Figure \ref{fig:dist},
the red spots present $33$ particles,
whose energy exceeds $90 mc^2$ at $t/\tau_{c}=100.0$.
They are found only near the magnetic pileup regions 
along the flat current sheet layer.
According to the partial and the main energy spectra
at $t/\tau_{c}=100.0$ in Figure \ref{fig:recspec},
the intermediate-energy particles ($40 mc^2 < \varepsilon < 70 mc^2$)
are found both inside and outside the selected region and
their spectral slopes look similar.
The selected region has fewer intermediate-energy particles by a factor of $3$,
because the region is narrower than the typical current sheet length.
On the contrary, high-energy particles
whose energy exceeds $\varepsilon > 80 mc^2$
are only found outside the selected region.
Therefore, these highest energy particles are accelerated
outside the selected region of $-12.8<x/\lambda<12.8$.
In fact, during $t/\tau_c=60.0$ and $80.0$,
the highest-energy shoulder increases
by $\Delta\varepsilon \sim (25-30) mc^2$,
and some particles apparently gain more energy than
the maximum energy gain inside the AR,
$\Delta\varepsilon \sim ( e E_y c \cdot 20 \tau_c ) \sim 25 mc^2$,
where we use $E_y\leq0.2B_0$ as the typical value. 
Then, we traced the most energetic positron,
whose energy is $97 mc^2$ at $t/\tau_{c}=100.0$
and who is found at $(x/\lambda,z/\lambda)=(31.0,-0.64)$.
Its trajectory and $xz$-/$xy$- projections
are presented in Figure \ref{fig:orbit}.
Labels in the $xy$ plane indicate the relevant times.
Time histories of the relevant properties are presented
in Figure \ref{fig:orbit2}, too.
Originally, the positron's motion is
well explained by the meandering motion;
it bounces around the neutral sheet in the $\pm z$ direction.
Around $t/\tau_c=(40\sim 50)$, the reconnection breaks up and
the positron enters the $X$-type region ($x/\lambda \sim 0$).
Then, it is driven by the reconnection electric field $E_y$
and its energy starts to increase.
In the gray background regions,
the field satisfies the condition of the AR, $(E^2-B^2) > 0$.
This particle does not always stay in the AR,
because it travels near the edge of the AR.
After $t/\tau_c=65$,
the positron departs from the central AR
and then it starts to travel through the (relativistic) Speiser orbit.
The energy already reaches up to $\varepsilon =50 mc^2$
through particle acceleration in/near the central AR.
At this stage, the positron's Larmor radius
is $(\gamma mc^2/eB_0) \sim 8 \lambda$. 
Because of a relativistic inertia effect,
it can travel a longer distance into the $y$-direction
through Speiser orbit.
If we observe this particle
in the moving frame of the plasma outflow,
its energy stays nearly constant after $t/\tau_c\gtrsim 65$.
In the simulation frame,
the particle gains further energy from $E_y$
until it is ejected into the $x$-direction.

\begin{figure}
\begin{center}
\includegraphics[width=\columnwidth,clip]{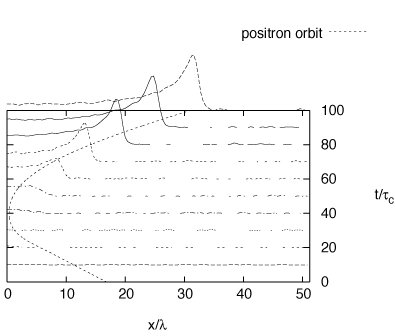}
\caption{$x$-$t$ diagram of the positron trajectory and the reconnection electric field $E_y$ along the neutral sheet ($z=0$).
\label{fig:stack}}
\end{center}
\end{figure}

Interestingly, in this case
the particle gains more energy
than the standard Speiser orbit case,
because the electric field $E_y$ increases in time
at the particle's position. 
Figure \ref{fig:stack} is an $x$-$t$ diagram of
the positron's position and the $E_y$-profile.
The dashed line presents the positron trajectory, and
the stack plots show the time development of
the reconnection electric field $E_y$ along the neutral sheet ($z=0$).
Surprisingly, the positron always stays around
the peak of the electric field $E_y$,
which is related to the magnetic pile-up region.
Since magnetic field lines are continuously transported
from the $X$-type region,
the piled-up peak increases in its height,
and then $E_y$ is further enhanced by
the relevant motional electric field. 
In other words, the particle speed resonates with
the propagating speed of the reconnection
pileup front $\sim 0.6-0.7c$,
which is slightly slower than the reconnection outflow speed.
In the geophysical ion-electron contexts,
\citet{birn97} studied ion acceleration and
\citet{hoshi01} studied electron acceleration
in the magnetic dipole/pileup regions.
In both cases, particles gain their energy from $E_y$,
through $\grad \vec{B}$ drift motion in the $y$-direction. 
In our case, the particle directly resonates with the fields. 
In order to stay for a long time in the piled-up front,
the particle has to be heavy enough
or the particle's Larmor radius has to be large enough.
Therefore, only high-energy particles
can be continuously accelerated to the highest energy;
most of them are preaccelerated around the AR
when they enter the piled-up regions.
Low-energy particles are quickly
ejected into the outflow directions. 
When these pileup regions hit the downstream dense regions,
it releases a lot of energy into the downstream plasma energy.
The downstream plasmas increase their energy
on the order of
$\Delta \varepsilon \sim 2 r'_L e E'_{y} \sim 3 (\gamma mc^2/e B_0) e B_0 \sim 3 \gamma mc^2$,
where the prime denotes the physical properties in the piled-up front frame,
and this mirror effect explains the mild enhancement of
the global energy spectra around $\varepsilon \sim (20-30) mc^2$.
If we compare the piled-up region with the AR,
the piled-up region mainly contributes to the plasma heating,
but it also enhances the maximum energy of a small number of
highly accelerated particles. 
A similar mechanism is reported in \citet{claus04},
although their difference between
the $dc$ acceleration in the centeral AR
and this piled-up acceleration was not clear.
Their system becomes highly turbulent,
because the main reconnection is suppressed by
the conductive wall boundary effect in the inflow region.
On the contrary, in our case,
the main reconnection is so powerful that
it sweeps out everything into the $\pm x$ direction,
including small tearing islands in its early stage.
Therefore, the acceleration continues for a longer time,
until outflow jets reach the periodic boundaries.
We do not know how long the single reconnection region can dominate,
but we remark that the higher maximum energy will be obtained
when a single reconnection dominates. 

\begin{figure}[htbp]
\begin{center}
\includegraphics[width=\columnwidth,clip]{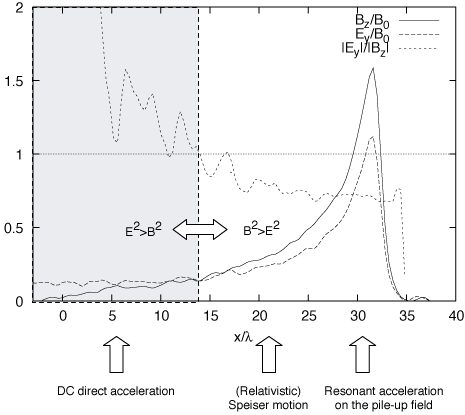}
\caption{
Field structure in the neutral sheet at $t/\tau_c=100.0$.
The normalized electromagnetic fields $B_z$ and $E_y$ and the ratio of $|E_y|/|B_z|$ are presented.
\label{fig:stack2}}
\end{center}
\end{figure}

Figure \ref{fig:stack2} summarizes the acceleration mechanism,
presenting the electromagnetic field properties
at $t/\tau_c=100.0$ along the neutral plane $z=0$.
As stated, the electric field is
larger than the magnetic field $B_z$
near the $X$-type region $x/\lambda < 14$,
while $B_z$ exceeds $E_y$ in $x/\lambda >14$.
The dashed line presents the ratio of $E_y$ to $B_z$,
which agrees with the plasma outflow speed
around the outflow the region.
In this figure, we can classify the acceleration site
into the following three types:

Central acceleration region ---Where $(E^2-B^2)>0$ and $dc$ direct acceleration takes place.

Outside region ---Where $(E^2-B^2)<0$ and particles are accelerated through the (relativistic) Speiser orbit.

Magnetic pileup region ---Where some high-energy particles resonate with the enhanced electric field.

Regarding the boundary effects on the late-time development,
the plasma collision across the $x$-boundaries
enhances the magnetic pileup flux and the relevant acceleration.
However, such a collision is quite likely to occur
in realistic situations.  For example,
the coalescence of multiple reconnection islands
have recently attracted our attention
because it generates energetic particles \citep{saito}.
Since our system width ($L_x =102.4\lambda$) is
rather larger than the typical scale of the tearing mode
($\sim 20\lambda$, assuming that the fastest mode is $k_x\lambda\sim0.3$),
we believe that the $x$-boundary effect is plausible.
The $z$-boundary effect may be rather artificial; however,
it is unlikely that the $z$-boundary effect accelerates
the reconnection process in the early stage of $t/\tau_c\lesssim 100$.
The inflow magnetic flux continually decreases in time
(approximately half of the original flux $\sim 0.5 B_0$
at $t/\tau_c=100.0$)
and then it starts to increase
in the later stages at $t/\tau_c\gtrsim 170$,
due to the $O$-point expansion
in the sub simulation domain.
In the striped pulsar wind condition,
it is possible that double/multiple current sheets
are close to each other.
Since we note that the late-time energy spectra become
even harder (with an index of $\sim -2.4$),
these results imply that
harder energy spectra may be generated
when the pulsar wind is crowded by the multiple current sheets.
Of course, it is very difficult to discuss the spectral index
from the simulation results,
because it shows the very early development in a small region
compared with actual astrophysical phenomena.
It just gives us a hint to discuss
an astronomical long-time/large-scale evolution
and the resulting energy spectra.

\subsection{Case studies}

We have further carried out five simulation runs
with different temperature parameters;
$T/mc^2=16, 4, 1, 1/4$ and $1/16$.
We call them runs R1-R5 and
their parameters are presented
in Table \ref{table}.
The background plasma temperature
$T_{bg}$ is always set to $0.1T$ and
$n_{bg}/(\gamma_{\beta}n_0)$,
the ratio of the number density in the simulation frame,
is fixed to $0.05$.
The amplitude and duration of triggering fields are
selected case by case, but we usually use smaller values than run R3.
For comparison, our previous work (ZH01) is presented as run R0,
whose parameters are equivalent to $T/mc^2 \sim 1/4$ and $\beta \sim 0.27$.

\begin{figure}[htbp]
\begin{center}
\includegraphics[width=\columnwidth,clip]{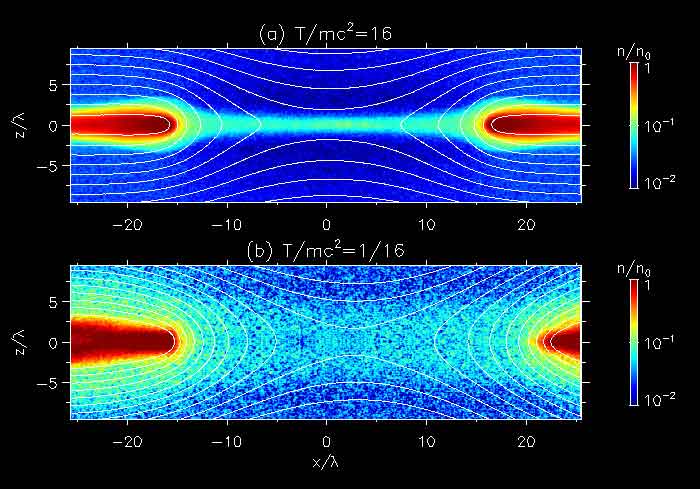}
\caption{\label{fig:rec4}
Snapshots of runs for $T/mc^2=16$ (\textit{top}) and $T/mc^2=1/16$ (\textit{bottom}).}
\end{center}
\end{figure}

\begin{figure}[htbp]
\begin{center}
\includegraphics[width=0.7\columnwidth,clip]{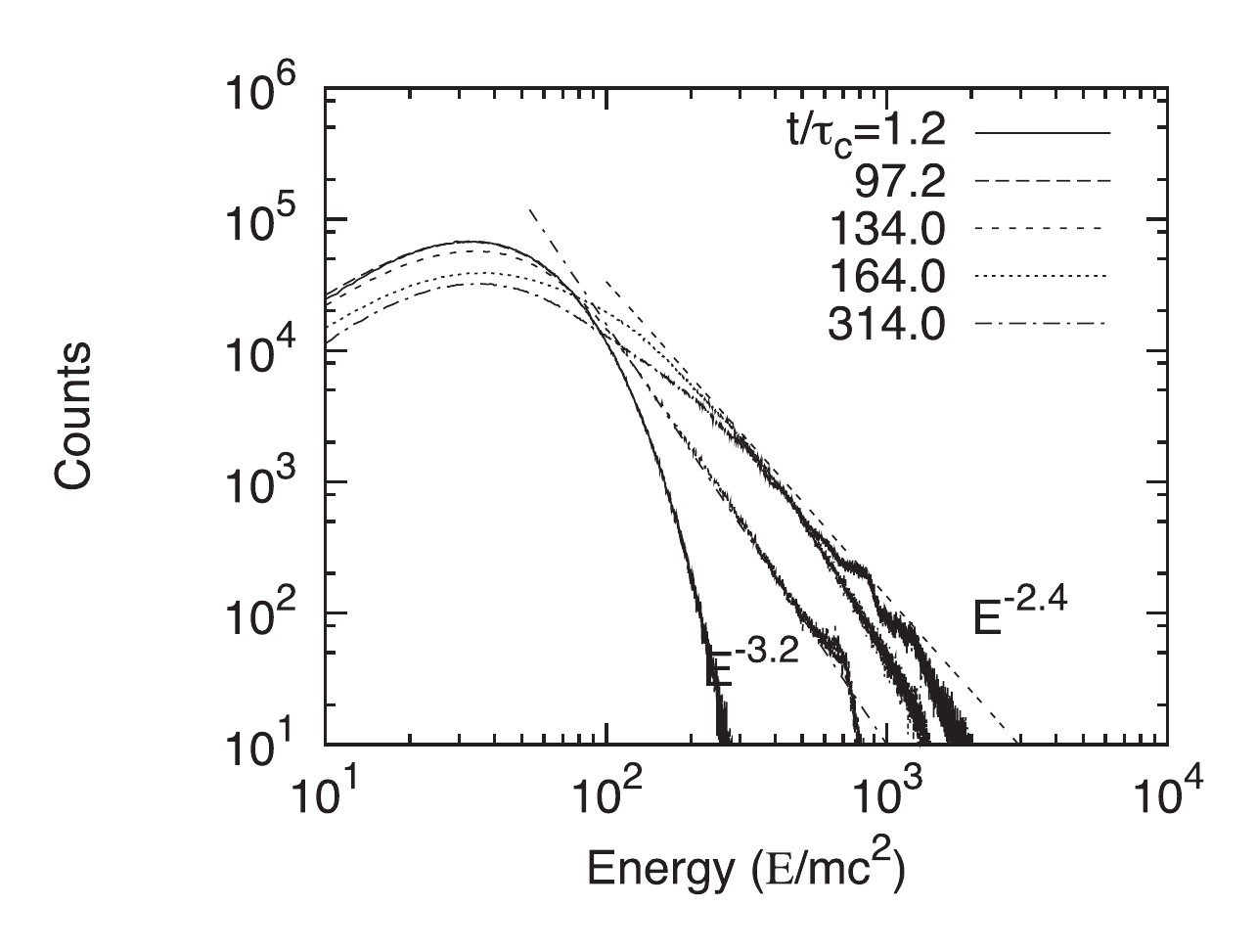}
\caption{\label{fig:recspecH}
Particle energy spectra in run R1 in log-log format.}
\end{center}
\end{figure}

\begin{figure}
\begin{center}
\includegraphics[width=0.95\columnwidth,clip]{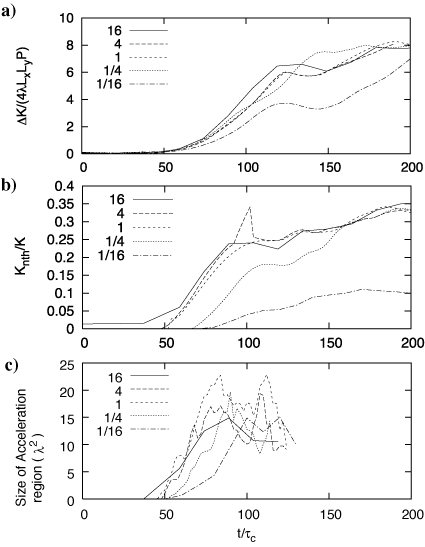}
\caption{\label{fig:rec_hist} \textit{(a)} Enhancements of the particle kinetic energies in the reconnection runs with $T/mc^2=16, 4, 1, 1/4, 1/16$.
Energies and time are normalized by the pressure in the original current sheet and
the light transit time, respectively.
\textit{(b)} Nonthermal ratio parameters for reconnection.
\textit{(c)} Size of the acceleration region.
}
\end{center}
\end{figure}

The results of the parameter studies are presented
in Figures \ref{fig:rec4}, \ref{fig:recspecH}, and \ref{fig:rec_hist} and 
Table \ref{tab:rec_alfven}.
Figure \ref{fig:rec4} shows typical snapshots
in runs R1 and R5 ($T/mc^2=16$ and $1/16$).
The energy spectra in run R1 are also presented in Figure \ref{fig:recspecH}.
The enhancements of plasma kinetic energies in runs R1-R5
are presented in Figure \ref{fig:rec_hist}\textit{a}.
Their energies are normalized by the plasma pressure energy
in the original current sheet,
which is roughly proportional to the temperature $T$.
The time axis is re-arranged so that
reconnection breaks up nearly at the same time.
In addition, in order to evaluate the acceleration amplitude,
we calculate the ``nonthermal ratio'' index of $K_{nth}/K$
in Figure \ref{fig:rec_hist}\textit{b},
where $K$ is the total kinetic energy in the main simulation box
and $K_{nth}$ is its nonthermal part.
We estimate $K_{nth}$ by comparing
the energy spectra and its equivalent ``thermal'' distribution.
Its derivation is described in Appendix A. 
Figure \ref{fig:rec_hist}\textit{c} presents
the size of the acceleration region
in the main simulation box,
which satisfies the condition of $(E^2-B^2)>0$.
Table \ref{tab:rec_alfven} shows characteristic speeds in the system.
We pick up the maximum plasma speed along outflow/inflow regions,
as long as it does not violate the MHD frozen-in condition so much.
The typical \Alfven velocity is calculated from the background magnetic field $B_0$
and plasma parameters in the current sheet $(n_{0},T)$.
The inflow \Alfven velocity is the local \Alfven speed
in the background region.

System evolutions are quite similar
in relativistic cases of $T/mc^2 \gtrsim 1$.
In run R1, the spatial structure (Fig. \ref{fig:rec4}\textit{a})
and the power-law index of $-3.2$ (Fig. \ref{fig:recspecH})
in the reconnection phase
are in good agreement with those of run R3.
In the very late time stage (at $t/\tau_c=314.0$),
the power-law index is approximately $-2.4$.
The evolution timescale of $(30-50) \tau_c$
and the saturation level (Fig. \ref{fig:rec_hist}\textit{a})
and a portion of the nonthermal energy (Fig. \ref{fig:rec_hist}\textit{b})
are almost the same, too.
The ratio of the inflow speed and the outflow speed
usually stay around $0.3$-$0.4$, except for run R1 (Table \ref{tab:rec_alfven}).
In run R1, the rate becomes slightly higher
due to the faster inflow speed.
However, it still stays nonrelativistic;
the relevant Lorentz factor is only $\sim 1.25$,
and the typical inflow speed is slower
because we picked up the highest inflow speed along the $y$ axis.
Considering that the system is magnetically dominated
$[(B_0^2/8\pi)/(\sum \varepsilon) \sim 60]$
in the inflow region and that
the reconnection layer's aspect ratio is $L/\delta \sim 10$,
where $L$ and $\delta$ are the typical width and height
of Sweet-Parker reconnection layer,
our results do not agree with \citet{lyutikov03};
they predicted that the inflow velocity becomes relativistic and
that the outflow velocity is faster than the inflow \Alfven speed
when the inflow region is magnetically dominated.

In less relativistic cases,
the current sheet structure is ambiguous or diffusive (Fig. \ref{fig:rec4}\textit{b}).
This is due to relatively large Larmor radius
(eq. [\ref{eq:gyro}]; $r_L \sim 0.8 \lambda$ in run R5)
in the nonrelativistic limit.
In addition, the system evolves slowly because of
the slower characteristic speed: the nonrelativistic \Alfven speed.
Contrary to relativistic MHD simulations \citep{naoyuki06},
in which the outflow speed is
well-approximated by the inflow \Alfven speed,
the maximum outflow speed stays between the two \Alfven speeds.
The size of the AR decreases
and then fewer nonthermal particles are observed.
Strictly speaking, this may contain ARs
in several $X$-points or magnetic nulls inside the $O$-points.
However, in the breakup stage of reconnection ($50 < t/\tau_c < 100$),
they represent the size of the AR in the single reconnection region very well.
We believe that the size of the AR is
controlled by the reconnection outflow speed.
The magnetic topology of reconnection is
rather insensitive to parameters
and so the relative size of the AR depends on
the amplitude of the reconnection electric field $E_y$.
Then, $E_y$ is controlled by the typical outflow speed
through the frozen-in condition at the outflow region,
$|v_{out}|=c|E_y|/|B_{z0}|$,
where $B_{z0}$ is the typical reconnected magnetic field.
In summary, in the relativistic regime,
the AR (Fig. \ref{fig:rec_hist}\textit{c}) becomes larger 
due to its relativistic outflow speed $v_{out}/c$,
and then reconnection generates
more nonthermal particles (Fig. \ref{fig:rec_hist}\textit{b}).

The direct acceleration around the AR is not likely to occur
in the solar flare regions or the Earth's magnetotail,
where the reconnection outflow is on the order of the \Alfven speed of
$10^6$ (m/s) $\ll c \sim 3 \times 10^8$ (m/s) and
the AR is restricted within the electron inertial length.

\section{Relativistic drift kink instability}

\subsection{Simulation result}

Next, we introduce the other two-dimensional simulations in the $y$-$z$ plane.
The system size is now set to
1 ($x$) $\times$ 256 ($y$) $\times$ 512 ($z$).
We set no artificial electric fields to excite instabilities,
since instabilities spontaneously arise from thermal noise.
We set no background plasmas.
In the first simulation, we set $T = mc^2$.
This run is run D3 in Table \ref{table}.

\begin{figure}[htbp]
\begin{center}
\includegraphics[width=\columnwidth,clip]{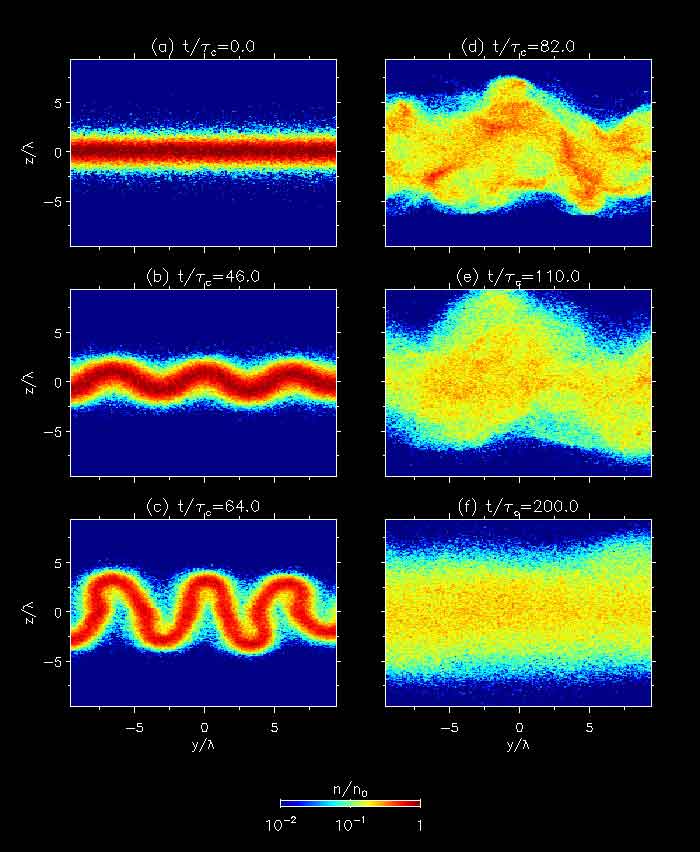}
\caption{\label{fig:kink1}
Snapshots of run D3 in the $yz$ plane at various stages ($t/\tau_c=0.0, 46.0, 64.0, 82.0, 110.0$, and $200.0$).}
\end{center}
\end{figure}

\begin{figure}[htbp]
\begin{center}
\includegraphics[width=0.7\columnwidth,clip]{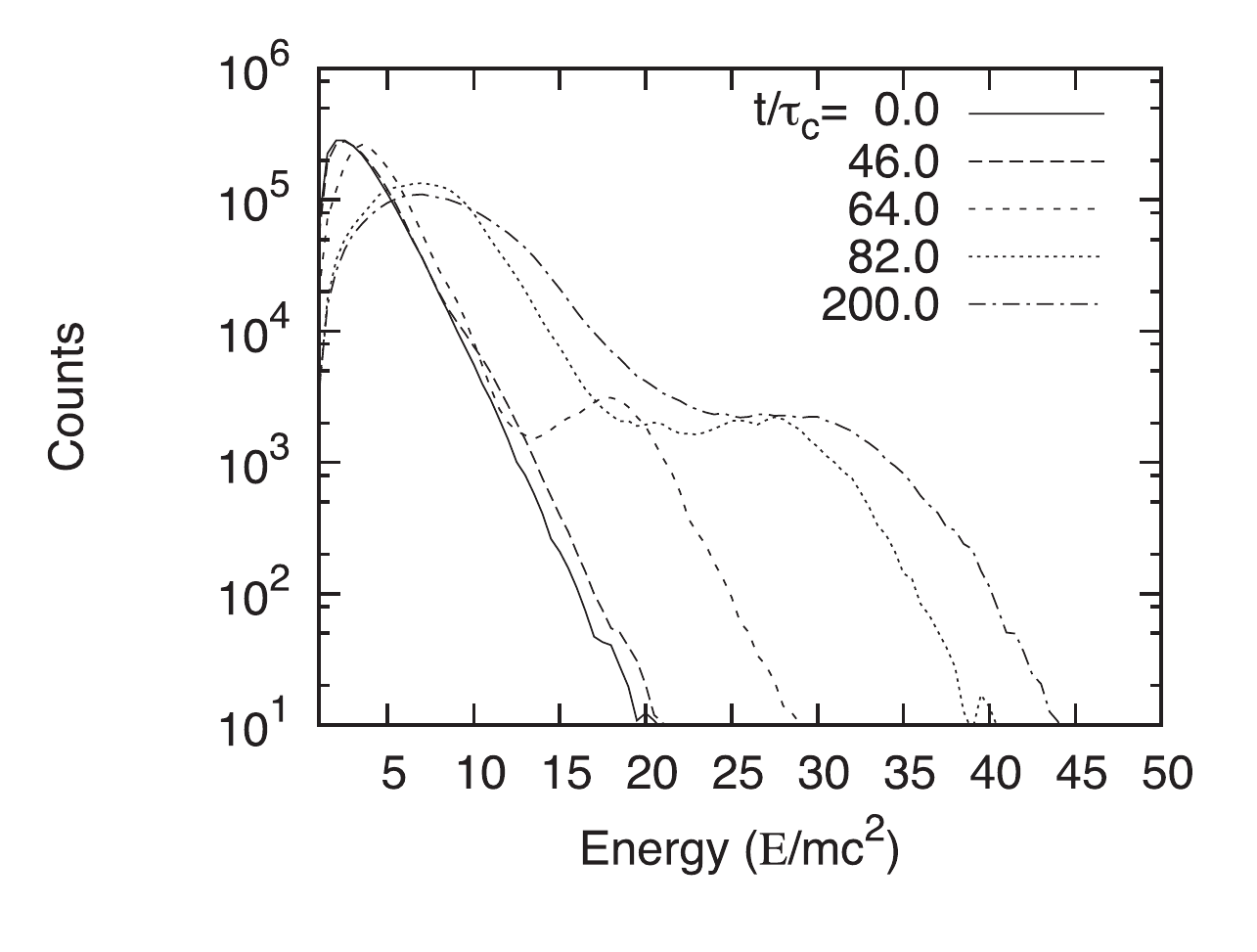}
\caption{\label{fig:kink2}
Energy spectra in run D3 at various stages.
The particle count number,
integrated over the main simulation box,
is presented as a function of
the particle energy.}
\end{center}
\end{figure}

\begin{figure}[htbp]
\begin{center}
\includegraphics[width=\columnwidth,clip]{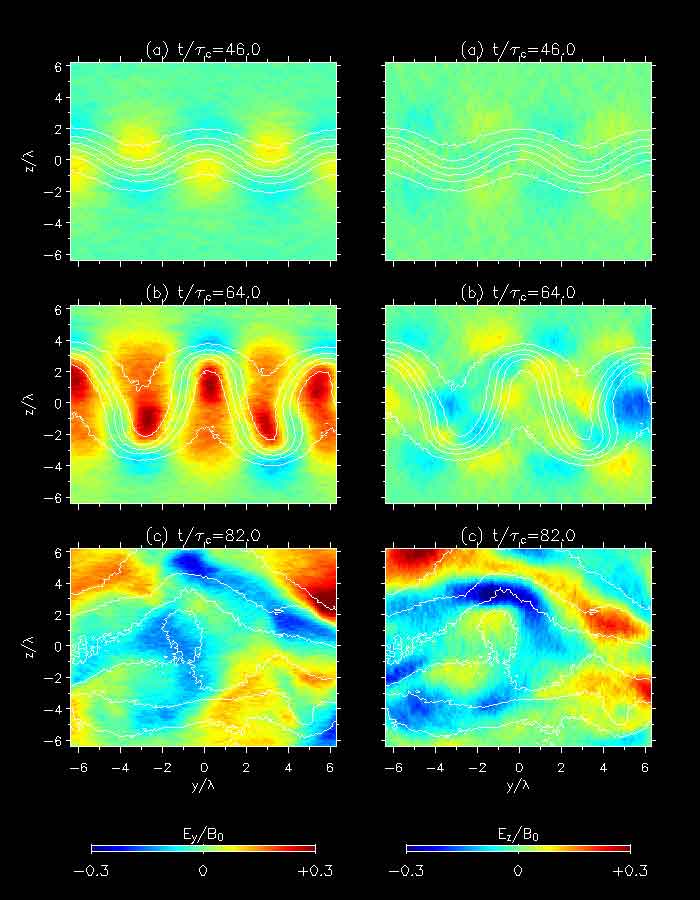}
\caption{\label{fig:kink3}
Electric fields $E_y$ and $E_z$ at $t/\tau_c=46.0, 64.0$, and $82.0$.
White lines show contours of $B_x$.
}
\end{center}
\end{figure}

Characteristic snapshots of the simulation
are presented in Figure \ref{fig:kink1}. 
The color contours shows the plasma density,
which is normalized by the peak density of
the Harris current sheet $\rho_0 = 2\gamma_{\beta}n_0$.
Figure \ref{fig:kink2} presents
the energy spectra at the corresponding times. 
Figure \ref{fig:kink3} presents
the electric structures of selected stages.
The left three panels of Figure \ref{fig:kink3}
are color contours of the electric field $E_y$.
The sign of $E_y$ is positive in light gray (yellow) regions,
while it is negative in dark gray (blue) regions.
The right panels of Figure \ref{fig:kink3}
show the electric field $E_z$ at the corresponding time.
White lines are contours of $B_x$,
whose structure is similar to the density profile.

After tens of the light transit time, we observe
a linear growth of a kink-type structure around $t/\tau_c=46.0$.
This is due to the relativistic drift kink instability (RDKI)
\citep{zeni05a},
a relativistic extension of the drift kink instability (DKI),
which is a long-wavelength,
current-driven instability in a thin current sheet.
In the top panels in Figure \ref{fig:kink3},
one can see a typical structure of the polarization electric fields
of $E_y$ and $E_z$ at this stage.
The $E_y$ plot in the top left panel shows
two types of color regions along the current sheet:
yellow means $E_y>0$ or in the right direction and
blue means $E_y<0$ or in the left direction.
They are anti-symmetric with the neutral plane ($z=0$).
Since $B_x$ is positive on the upside of the current sheet
and negative on the downside of the current sheet,
the observed sign of $\vec{E} \times \vec{B}$ is 
consistent with the $z$-displacement of the plasma bulk motion.
The other components of electromagnetic fields,
$E_x$, $B_y$, and $B_z$, are negligible. 
After $t/\tau_{c} \sim 50$,
the system turns into its nonlinear stage.
The typical signature is presented at $t/\tau_{c} = 64.0$
in Figures \ref{fig:kink1}, \ref{fig:kink2}, and \ref{fig:kink3}.
In this stage, the current sheet is strongly modulated
by the RDKI around $-4 < z/\lambda < 4$.
Because of the $z$-displacement by the RDKI,
we see the characteristic regions in a row along neutral plane ($z = 0$)
in the middle left panel of Figure \ref{fig:kink3}.
Originally, these regions were alternately arranged
on the upper side and on the lower side of the neutral plane
in the top left panel of Figure \ref{fig:kink3}.
Importantly, we observe a clear sign of the particle acceleration
in high-energy tails in the energy spectra in Figure \ref{fig:kink2}.
We discuss the acceleration process later.

After the nonlinear development,
the folded current sheets start to collapse each other,
and then the system turns into the ``mixed'' stage.
Typical signatures are presented in snapshots at $t/\tau_{c} = 82.0$
in Figures \ref{fig:kink1}, \ref{fig:kink2} and \ref{fig:kink3}.
Particle acceleration seems to stop
because the acceleration channel structure disappears.
Because of the collisions of sheet fragments,
magnetic energy is diffused into plasma heat,
and then the total kinetic energy has increased by 170\% at this time.
Finally, the mixed current sheet slowly evolves into
the broadened current sheet,
which is 3 or 4 times thicker than the initial sheet width.
The signature of this stage is presented in the snapshots
at $t/\tau_c=200.0$ in Figures \ref{fig:kink1} and \ref{fig:kink2}.
The spectrum looks unchanged from the mixed stage;
one can still see the remnant of the nonthermal tail.
We still recognize both
a remnant of the nonthermal tail and heated plasma component
in the spectrum at $t/\tau=200.0$ in Figure \ref{fig:kink2}.
Throughout Run D3, the total energy is
conserved within an error of $0.3\%$.

\subsection{Linear analysis}

\begin{figure}[htbp]
\begin{center}
\includegraphics[width={0.7\columnwidth},clip]{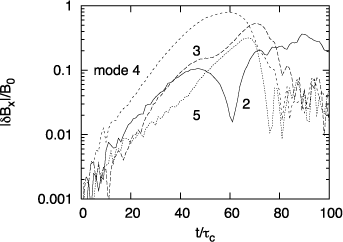}
\caption{Time histories of the amplitude of perturbed magnetic field $\delta B_x$
in the neutral plane ($z=0$)
for drift kink modes. Modes 2-5 correspond to
$k_y\lambda=0.49,0.74,0.98,$ and $1.22$.
\label{fig:mode}}
\end{center}
\end{figure}

Throughout the simulation run,
we have observed the perturbed magnetic field $\delta B_x$
in the neutral plane ($z=0$).
We have examined the periodic perturbation modes and
their time histories are presented in Figure \ref{fig:mode}.
Each mode ($2$-$5$) corresponds
to $k_y\lambda=0.49,0.74,0.98$, and $1.22$, respectively.
The most dominant mode is mode 4 with $k_{y}\lambda = 0.98$.
Its linear growth rate is
$\omega_i \sim 0.11 \tau_c^{-1}$ or $\omega_i \sim 0.035\Omega_c$,
where $\Omega_c=\omega_c/\gamma=(eB)/(\gamma mc^2)$ is
the typical gyro frequency.

\begin{figure}[htbp]
\begin{center}
\includegraphics[width={\columnwidth},clip]{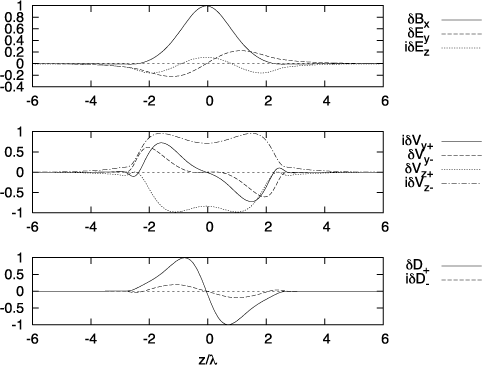}
\caption{The $z$-profiles of the eigen functions of the relativistic drift kink instability (RDKI) for the parameter $k_y \lambda = 1.0$.
\textit{Top}: Perturbations of $\delta B_x$, $\delta E_y$ and $\delta E_z$.
\textit{Middle}: Perturbed plasma velocities.
The bulk velocities of plasmas are $\delta V_{\{y,z\}+} = \delta v_{\{y,z\}p} + \delta v_{\{y,z\}e}$ and
the relative velocities of the two species are $\delta V_{\{y,z\}-} = \delta v_{\{y,z\}p} - \delta v_{\{y,z\}e}$.
\textit{Bottom}: The density perturbations are presented: $\delta D_+ = \delta d_p + \delta d_e $ and $\delta D_- = \delta d_p - \delta d_e$.
\label{fig:eigen}}
\end{center}
\end{figure}

\begin{figure}
\begin{center}
\includegraphics[width={\columnwidth},clip]{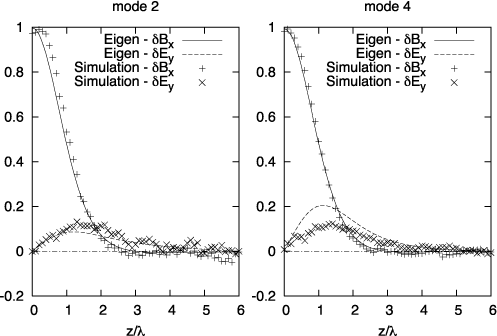}
\caption{Perturbation profiles of
the simulation data and relevant eigen modes.
\textit{Left}: Mode 2 ($k\lambda=0.49$) and the eigenmode for $k\lambda=0.50$.
\textit{Right}: Mode 4 ($k\lambda=0.98$) and the eigenmode for $k\lambda=1.00$.
\label{fig:dkisim}}

\includegraphics[width={0.8\columnwidth},clip]{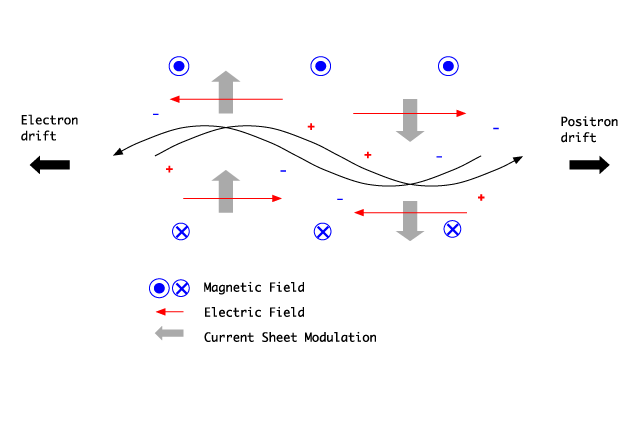}
\caption{
Schematic explanation of charge separation in the RDKI
\label{fig:dki_logic}}
\end{center}
\end{figure}

\citet{prit96} (hereafter P96) analyzed the linear stability of the DKI
with fluid equations and Maxwell equations.
We employed a similar method to our case
with the relativistic fluid equation
\begin{equation}
\frac{\gamma^2_s}{c^2}
(p_s + \varepsilon_s)
( \pp{}{t} + \vec{v}_s \cdot \grad )
\vec{v_s}
=
- \grad p_s +
\gamma_s q_s n_s 
\Big( \vec{E} +  \frac{\vec{v}_s}{c} \times \vec{B}  \Big)
- \frac{\vec{v}_s}{c^2} ( \gamma_s q_s n_s \vec{E} \cdot \vec{v}_s + \pp{p_s}{t} ),
\label{eq:rfluid}
\end{equation}
the particle conservation law
\begin{equation}
\pp{}{t}(\gamma_s n_s) + \div(\gamma_s n_s \vec{v}_s ) = 0,
\end{equation}
and adiabatic gas condition with polytropic index $\Gamma = 4/3$
\begin{equation}\label{eq:P_ad}
p_s \propto n_s^{\Gamma}.
\end{equation}
The polytropic index $\Gamma$ should go down to $5/3$ in the nonrelativistic limit;
however, setting $\Gamma=5/3$ provides no significant change in our calculation.
The detailed formulation is presented in Appendix B.
We have to be careful because
the typical scale of particle motions are
comparable to the plasma sheet thickness:
the gyroradius $c/(eB/\gamma mc) \sim 0.3\lambda$ or
the meandering width
$[ (c/\Omega_c) \lambda ]^{-1/2} \sim 0.6 \lambda$.
Therefore, fluid theory and the adiabatic gas condition
are no longer valid in small scale structures. 
In addition, the theory is less trustable
when we choose a larger $\beta$ parameter,
because the typical gyro scale becomes larger (eq. [\ref{eq:rgyro}]).
The eigenprofiles for $k_y\lambda = 1.0$,
which is close to
the most dominant mode (mode 4; $k_y\lambda = 0.98$) in run D3,
are presented in Figure \ref{fig:eigen}.
In the top panel in Figure \ref{fig:eigen},
perturbations of the magnetic field ($\delta B_x$) and
the electric fields ($\delta E_y$ and $\delta E_z$)
are shown as functions of $z$.
They are normalized by the maximum amplitude of $\delta B_x$.
The anti-symmetric structure of $\delta E_y$ corresponds to
the positive and negative $E_y$ regions
along the neutral plane ($z=0$).
The other three components
($\delta B_y, \delta B_z, \delta E_x$) are zero in this case.
The $z$-profiles of velocity perturbations are
presented in the middle panel in Figure \ref{fig:eigen}.
The parameters
$\delta V_{y\pm} = \delta v_{yp} \pm \delta v_{ye}$
and
$\delta V_{z\pm} = \delta v_{zp} \pm \delta v_{ze}$
show the bulk or relative motion of the two species.
The $x$ components of the velocity perturbations ($\delta v_{x\{p,e\}}$) are zero.
The plasma bulk motion in the $z$-direction,
$\delta V_{z+} = \delta v_{zp} + \delta v_{ze}$,
is dominant around the neutral plane.
The relative $z$-motion $ V_{z-} = \delta v_{zp} - \delta v_{ze}$ is also dominant around the neutral plane.
It is mainly due to $z$-projection of
the original counter-streaming current.
A small-scale structure around $|z/\lambda| \sim 2-3$
may be numerical noise, which we could not completely clear out. 
The bottom panel in Figure \ref{fig:eigen} shows
the density perturbation of $\delta D_{\pm} = \delta d_p \pm \delta d_e$,
where $d=\gamma n$ is the plasma density in the observation frame.
The anti-symmetric structure of $\delta D_{+}$ is consistent with
the $z$-displacement of the central current sheet. 
Generally, their structures are
in good agreement with the obtained simulation data,
except for unphysical small structures
(i.e. small waves in $i\delta V_{y+}$ around $z/\lambda=\pm 2.5$).
The perturbation profiles in the simulation at $t/\tau_c=40.0$
and the eigen functions are presented in
Figure \ref{fig:dkisim}.
The left panel compares mode 2 ($k\lambda=0.49$)
in simulation and eigen mode for $k\lambda=0.50$.
Both are in excellent agreement,
although $\delta E_y$ in simulation looks slightly noisy
because of its weak amplitude.
The right panel of Figure \ref{fig:dkisim}
compares mode 4 ($k\lambda=0.98$)
and eigen mode for $k\lambda=1.00$.
They are in good agreement in $\delta B_x$,
while $\delta E_y$ in simulation is
approximately half of the eigen profile.
The difference is also found in their the growth rates;
the eigen growth rate at $k\lambda \sim 1.0$ is $\omega_i=0.19 \tau_c^{-1}$.
We believe the fluid instability is dumped by the kinetic effect,
or, in other words, fluid theory is losing its validity
around the short wavelength of $k\lambda \gtrsim 1$.
We discuss the validity of our theory later again.
The charge separation ($D_{-}$) is schematically described in Figure \ref{fig:dki_logic}.
Once the current sheet is modulated,
the streaming $e^{+}$ fluid and the counter-streaming $e^{-}$ fluid
are slightly separated in $\pm x$ direction,
due to the particle inertia or
the drift motion between the centrifugal force and magnetic fields.
Then the separated charges produce
the electrostatic fields $E_y$ and $E_z$.
Since the $E_y \times B_x$ motion enhances the current sheet modulation in the $z$ direction,
the instability continues to grow. 
There is no influence of the Kelvin-Helmholtz instability
in the system.
Since we set empty background plasmas,
there is no velocity shear between
the current sheet and the background region.
We also compare run D3 and the other run
with 5\% of background plasmas (run D3a in Table \ref{table}), and
we confirm that their growth rates are the same.

\subsection{Analytical theory}

Following P96, we construct a simple analytic solution of the instability
in a long-wavelength limit and with $\beta \ll 1$.
We drop the third term from the right-hand side of equation \eqref{eq:rfluid},
because they are smaller than other terms when linearized in a limit of $\beta \ll 1$.  We get
\begin{equation}
\frac{ (\vec{v}_s/{c^2}) ( \gamma_s q_s n_s \delta\vec{E} \cdot \vec{v}_s ) }{\gamma_s q_s n_s \delta\vec{E} } \sim O(\beta^2),
\frac{ (\vec{v}_s/{c^2}) \partial\delta p_s/{\partial t} }{\grad \delta p_s}
\sim 
O(\beta \frac{c_s}{c} )
\lesssim O(\frac{\beta}{\sqrt{3}}),
\end{equation}
where $c_s$ is the sound speed.
In addition, using the density ($n_s\gamma_s = d_s$) in the observer's frame
and rewriting the enthalpy term
$M=m\gamma_s ( 1 + [\Gamma/(\Gamma-1)] [T/({mc^2})] )$,
we obtain
\begin{equation}\label{eq:rfluid22}
Md_s
( \pp{}{t} + \vec{v}_s \cdot \grad )
\vec{v_s}
=
- \grad p_s +
d_s q_s
\Big( \vec{E} +  \frac{\vec{v}_s}{c} \times \vec{B}  \Big)
\end{equation}
and the equation of continuity
\begin{equation}\label{eq:rDcont}
\pp{d_s}{t} + \div( d_s \vec{v}) = 0.
\end{equation}

Using P96's method (described in the Appendix in P96)
in equations \eqref{eq:rfluid22} and \eqref{eq:rDcont} and the Maxwell equations,
one can obtain the modified growth rate of the DKI,
\begin{equation}
\gamma_{RDKI}/ ( k_y c\beta )
\approx
1 - 2
\Big( \frac{c\beta }{ (eB_0/Mc) \lambda }
\Big)
k_y \lambda
= 1 - 2 \gamma_{\beta} ( 1 + \frac{\Gamma}{\Gamma-1}\frac{T}{mc^2} )
\Big( \frac{c\beta }{ (eB_0/mc) \lambda }
\Big)
k_y \lambda,
\label{eq:rdki}
\end{equation}
where $\gamma_{RDKI}$ is the growth rate of the RDKI. 
Since equation \eqref{eq:rdki} gives larger results than
the actual growth rates in nonrelativistic studies \citep{prit96,dau99},
we assume that its right-hand side of the equation gives
an upper limit of the growth rate.
Using equations \eqref{eq:pb} and \eqref{eq:curr},
we obtain $\gamma_{RDKI}$ as a function of $k_y\lambda$,
\begin{equation}\label{eq:rdki2}
\tau_c \gamma_{RDKI} 
\approx
[
1 - \gamma_{\beta} \Big(
\frac{mc^2}{T} + 
\frac{\Gamma}{\Gamma - 1}
\Big)
\beta^2
(k_y \lambda)
]
\cdot
\beta
(k_y \lambda) .
\end{equation} 
This equation has a maximum value
\begin{eqnarray}\label{eq:dki4}
\tau_c \gamma_{RDKI} = \frac{1}{4\gamma_{\beta}\beta}
\Big( \frac{mc^2}{T} + \frac{\Gamma}{\Gamma-1}\Big)^{-1}
= \left\{ \begin{array}{ll}
(4\gamma_{\beta}\beta)^{-1}
({T}/{mc^2}) & (T \ll mc^2) \\
(16\gamma_{\beta}\beta)^{-1} & (T \gg mc^2) \\
\end{array} \right.
\end{eqnarray}
at
\begin{eqnarray}\label{eq:dki5}
k_y\lambda = \frac{1}{2 \gamma_{\beta} \beta^2}
\Big( \frac{mc^2}{T} + \frac{\Gamma}{\Gamma-1}\Big)^{-1}
= \left\{ \begin{array}{ll}
(2\gamma_{\beta}\beta^2)^{-1}
({T}/{mc^2}) & (T \ll mc^2) \\
(8\gamma_{\beta}\beta^2)^{-1} & (T \gg mc^2) \\
\end{array} \right.
~.
\end{eqnarray}
Thus, as long as $k_y \lambda < 1$ is satisfied in equation \eqref{eq:dki5},
equation \eqref{eq:dki4} gives a plausible upper limit of $\gamma_{RDKI}$.
In addition, equations \eqref{eq:dki4} and \eqref{eq:dki5} can be rewritten as
\begin{equation}\label{eq:upper}
\tau_c\gamma_{RDKI} < (k\lambda_y/2) \beta < \beta .
\end{equation}
When $k_y\lambda \gtrsim 1$ in equation \eqref{eq:dki5},
we use $k_y\lambda = 1$ in equation \eqref{eq:rdki2} to
obtain a plausible extension of the upper limit,
considering that the original equation \eqref{eq:rdki} is no longer valid in $k_y\lambda \gtrsim 1$.  We have
\begin{eqnarray}\label{eq:rdki5}
\tau_c \gamma_{RDKI} 
=
[
1 - \gamma_{\beta} \Big(
\frac{mc^2}{T} + 
\frac{\Gamma}{\Gamma - 1}
\Big)
\beta^2
]
\cdot
\beta
= \left\{ \begin{array}{ll}
[1 - \gamma_{\beta} \beta^2  ({mc^2}/{T})] \beta & (T \ll mc^2) \\
(1 - 4 \gamma_{\beta} \beta^2) \beta
 & (T \gg mc^2) \\
\end{array} \right.
~.
\end{eqnarray}
Note that both equations \eqref{eq:upper} and \eqref{eq:rdki5}
satisfy the simple condition
\begin{equation}\label{eq:rdki7}
\tau_c \gamma_{RDKI} < \beta .
\end{equation}

\begin{figure}[htbp]
\begin{center}
\includegraphics[width={\columnwidth},clip]{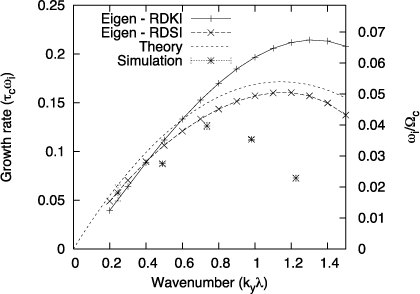}
\caption{Growth rates of the RDKI ($\omega_i$)
as a function of the normalized wavenumber ($k_y \lambda$).
Eigen growth rates (\textit{solid line with plus symbols}),
another branch of eigen growth rates (\textit{dashed line with crosses}),
eq. \eqref{eq:rdki2} (\textit{dashed line}),
and simulation results (\textit{asterisks}).
The timescale is presented in units of $\tau_c$ (\textit{left axis}) and
$\Omega_c$ (\textit{right axis}).
\label{fig:growth}}
\end{center}
\end{figure}

Figure \ref{fig:growth} compares the linear growth rate of the instability
in units of the light transit time of $\tau_c=\lambda/c$ (\textit{left axis})
or
the typical gyro frequency $\Omega_c$ (\textit{right axis}),
as a function of the normalized wavenumber ($k_y \lambda$).
The calculated eigen growth rates,
the analytical estimate (eq. [\ref{eq:rdki2}])
and the observed growth rates in the simulations are presented.
The observed growth rates correspond to
modes $2$-$5$ in Figure \ref{fig:mode}. 
We discuss the RDSI line in \S 4.4. 
The three values are in good agreement when $k_y\lambda \lesssim 0.7$.
In shorter wavelengths of $k_y \lambda \gtrsim 1$,
the simulation results do not agree with the theories
because the instability is suppressed by kinetic effects.
\cite{dau99} investigated the nonrelativistic DKI using Vlasov code, and
his result also agreed with the fluid theory,
when $k_y\lambda \lesssim 0.7$.

\subsection{Relativistic drift sausage instability}

\begin{figure}[htbp]
\begin{center}
\includegraphics[width={\columnwidth},clip]{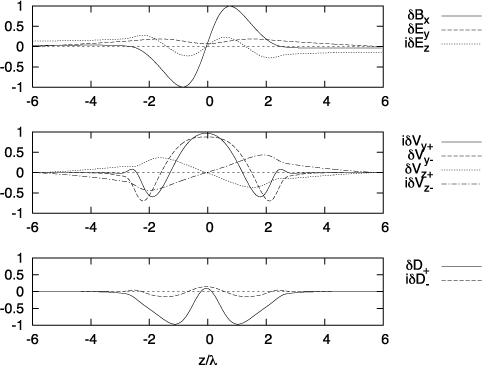}
\caption{Same as Fig. \ref{fig:eigen} but for the relativistic drift sausage instability (RDSI) with the parameter $k_y \lambda = 0.25$.
\label{fig:rdsi}}
\end{center}
\end{figure}

The RDSI line in Figure \ref{fig:growth}
stands for the relativistic extension of the drift sausage instability,
a cousin mode of the drift kink instability \citep{bu99,yoon01,silin03}.
Hereafter, we call it the relativistic drift sausage instability (RDSI).
Its eigen profiles at $k\lambda = 0.25$ are
presented in Figure \ref{fig:rdsi}.
The anti-symmetric profile of $\delta B_x$ and 
the symmetric profile of $\delta D_{+}$
are consistent with its sausage-type structure.
The RDSI is usually faster than the RDKI
in a very long wavelength limit of $k\lambda \ll 1$.
On the other hand, the fastest RDKI is usually faster than the RDSI.
In our runs, we could not observe any sign of the RDSI.
This result is consistent with P96, who suggested that
Landau damping stabilizes the (nonrelativistic) drift sausage instability.

\subsection{Particle acceleration}

\begin{figure}[htbp]
\begin{center}
\includegraphics[width=\columnwidth,clip]{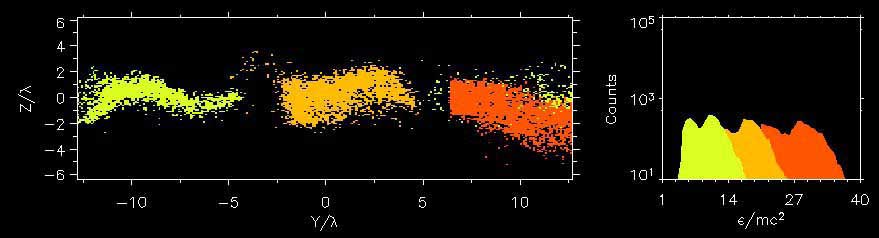}
\caption{\label{fig:dkidist}
Spatial distributions and energy spectra
of selected positrons, which satisfies the following two criteria:
(1) the kinetic energy satisfies $\varepsilon/mc^2 > 20$ at $t/\tau_c=$82.0
and (2) the $y$-position satisfies $6.4 \le y/\lambda < 12.8$ at $t/\tau_c=$82.0.
Three stages of the distribution are presented,
$t/\tau_c=46.0$ (light gray or yellow),
$64.0$ (gray or light orange) and
$82.0$ (dark gray or deep orange).
}
\end{center}
\end{figure}

In this section, we discuss
the particle acceleration in the nonlinear stage,
based on our previous works \citep{zeni05a}.
To study the acceleration site,
we select high-energy positrons
whose $y$-positions ($y$) and kinetic energies ($\varepsilon$) at $t/\tau_c=82.0$
satisfy the conditions $6.4 \le y/\lambda < 12.8$ and $\varepsilon/mc^2 > 20$.
Their spatial distributions and energy spectrum at three stages
(light gray at $t/\tau_c=46.0$, gray at $t/\tau_c=64.0$, and
dark gray at $t/\tau_c=82.0$) are presented in Figure \ref{fig:dkidist}.
The selected positrons are accelerated
in an ``acceleration channel'' (AC) around $z \sim 0$ into the $+x$ direction. 
Inside the AC, positive $E_y$ regions are in a row
(e.g. middle left panel in Fig. \ref{fig:kink3}),
and then high-energy particles successively skip across
the multiple positive $E_y$ regions along the AC,
once their Larmor radii are comparable with or
larger than the quarter wavelength of the RDKI.
In energy spectra at $t/\tau_c=46$ and $64$ in Figure \ref{fig:kink2},
particle acceleration effect is only observed
in the high-energy range of $\varepsilon \gtrsim 10-12 mc^2$,
where the gyro radius ($r_L=\gamma (c/\omega_c) \gtrsim (1.5-2) \lambda$)
exceeds the quarter wavelength $\pi/(2k_y) \sim 1.6 \lambda$. 
This threshold condition $\gamma c / \omega_c \gtrsim \pi / (2 k_y)$
yields
\begin{equation}
\varepsilon \gtrsim \pi eB_0 \lambda / (2 k_y \lambda) = \pi T / (\beta ~ k_y \lambda) \sim \pi T / \beta.
\label{eq:threshold}
\end{equation}
So, the particle acceleration in the AC will smear out
in the thick current sheet of $\beta \ll 1$,
because only a few particles satisfy this criterion.

The folded structure evolves into the mixed stage
when its $z$-displacement ($\Delta z$) is nearly the
same as the half-wavelength of the RDKI ($\Delta z \sim \pi/k_y\sim\pi\lambda$). 
So, in \citet{zeni05a}, we evaluated the timescale of
the nonlinear stage ($\tau_N$) as
\begin{equation}\label{eq:time_n}
\tau_{N} \sim \pi\lambda/\bar{v}_z \sim (\pi/\beta) \tau_c,
\end{equation}
where we assume that the typical $z$-displacement speed $\bar{v}_z \sim (0.2-0.3) c$ 
is approximated by the typical counter-streaming speed of $\beta c$.
After that, a lot of low-energy particles in the current sheet
start to interact with the electric fields.
The magnetic energy consumption is well explained by Joule heating
between the average electric field $\bar{E_y}$ and the zeroth order current
$\bar{J_y} \sim  cB_0 / (4\pi \Delta z)$ in the broadened current sheet.
We also approximated the dissipation timescale ($\tau_D$)
considering the energy consumption rate:
\begin{equation}
\tau_D 
\sim
( 2 \Delta z {B_0^2}/{8\pi} ) ( 2 \Delta z \bar{E_y} \bar{J_y} )^{-1}
\sim
( \pi / \beta ) \tau_c \label{eq:time_d},
\end{equation}
where $\bar{E_y} \sim (1/2)(\bar{v}_z/c) B_0$
and where the factor of $(1/2)$ represents that
the electric field is not uniform around the AC. 
From equations \eqref{eq:pb}, \eqref{eq:curr}, \eqref{eq:time_n},
and \eqref{eq:time_d},
we estimate the maximum energy gain
during the nonlinear stage and the mixed stage,
\begin{eqnarray}
\Delta \varepsilon_{est}
\sim
ec\bar{E_y} (\tau_N + \tau_D)
\sim
e B_0 \pi \lambda
\sim
2\pi T/ \beta . \label{eq:dki_max}
\end{eqnarray}
Typical estimated values
$\varepsilon_{est} = \varepsilon_{max0} + \Delta \varepsilon_{est}$
are presented in Table \ref{table}.

\subsection{Case studies}

\begin{figure}[htbp]
\begin{center}
\includegraphics[width=\columnwidth,clip]{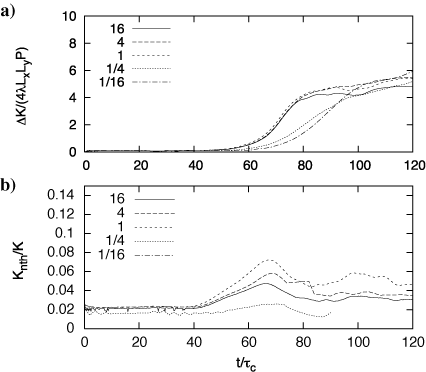}
\caption{\label{fig:dki_hist}
\textit{(a)} Enhancements of the particle kinetic energies in the RDKI runs.
Energies and time are normalized by the pressure in the original current sheet
and the light transit time, respectively.
\textit{(b)} Ratio of nonthermal energy in the case of the RDKI runs.
The ratio for $T/mc^2=1/16$ cannot be calculated by our method.
}
\end{center}
\end{figure}

We have further carried out simulation runs
with various temperatures,
runs D1-D5 ($T/mc^2=16,4,1, 1/4,$ and $ 1/16$) in Table \ref{table}.
We observe current sheet modulation and
the electromagnetic perturbations by the RDKI in all five cases.
Relativistic runs ($T/mc^2 \gtrsim 1$) show
similar results in many aspects of energy spectra,
wavelength, and growth rates. 
The energy spectra in runs D1 and D2 have
both the hot thermal component and the nonthermal component
and they look ``proportional'' to those in run D3.
The wavelength and the growth rates are nearly the same,
when normalized by the sheet thickness or the light transit time.
In less relativistic runs ($T/mc^2 < 1$),
we observe longer wavelength modes (mode 2; $k_y\lambda=0.49$)
as predicted by the theory (eq. [\ref{eq:dki5}]).
A high-energy tail becomes less apparent (Figure \ref{fig:dki_hist}\textit{b}) and
the RDKI grows slower than the theory (eq. [\ref{eq:dki4}]).
This is probably because the fluid theory loses validity 
in the nonrelativistic limit (eq. [\ref{eq:gyro}]).

Figure \ref{fig:dki_hist}\textit{a} shows
enhancements of the particle kinetic energies or
the released magnetic energy in runs D1-D5
with $T/mc^2=16, 4, 1, 1/4, 1/16$.
Energies are normalized by the pressure energy
in the original current sheet.
The saturation level always seems to be $5-6$, and
the timescale for the energy release is $t\sim 20 \tau_c$.
Nonrelativistic cases take more time due to their slower growth rates. 
Comparing with reconnection cases in Figure \ref{fig:rec_hist}\textit{a},
the amount of released energy is comparable; however,
due to its faster growth rates
their timescales are shorter in the relativistic regime. 
Figure \ref{fig:dki_hist}\textit{b} presents
the nonthermal ratio parameter of $K_{nth}/K$.
We could not resolve the nonthermal part with our method
in run D5 ($T/mc^2=1/16$) and
in the late-thermalized stage of run D4 ($T/mc^2=1/4$).
The parameter seems to become small as a smaller $T$ is used and
the particle acceleration is enhanced in the relativistic regime of $T/mc^2 \gtrsim 1$.
Comparing Figures \ref{fig:rec_hist}\textit{b} and \ref{fig:dki_hist}\textit{b},
apparently reconnection produces more nonthermal energy than the RDKI.
Although the width of the simulation box is not the same
($25.6 \lambda [y]$ for the RDKI, $102.4 \lambda [x]$ for reconnection),
the RDKI's nonthermal ratio will not change.
If we look at the smaller region around the $X$-type region,
we will obtain a higher nonthermal ratio.
Remember the partial spectrum at $t/\tau_c=100.0$ in Figure \ref{fig:recspec}.
Apparently, the partial spectra is highly dominated by nonthermal energy
and it is integrated over $-12.8<x/\lambda<12.8$---
the same width as the RDKI simulation domains ($25.6 \lambda$).
So, as an origin of the nonthermal particles,
reconnection is a more favorable candidate
due to the acceleration processes around the $X$-type region.
On the contrary, the RDKI quickly converts
magnetic energy into plasma thermal energy or plasma heat.

The maximum energies ($\varepsilon_{max}$) obtained by the RDKI
are presented in Table \ref{table}.
They are compared with our estimate from equation \eqref{eq:dki_max},
which seems to be a good approximation for these runs with $\beta=0.3$.
This relation will not be valid
in the ``thick'' current sheet with small $\beta$.
In thicker current sheets,
the Larmor radius becomes smaller compared with the wavelength,
so that particle acceleration may not take place. 
The maximum energies ($\varepsilon_{max}$) obtained by reconnection
are also presented in Table \ref{table}.
As far as we have investigated,
maximum energies by reconnection are far larger than those by the RDKI,
and reconnection is undoubtedly favorable to accelerate high-energy particles.
By the way, we do not know the upper limit energy of acceleration
by the single reconnection.
In Table \ref{table}, we roughly estimate the maximum energy by
$\varepsilon_{est}=eE_yc(L_x/2c)=eB_0L_x/2$
considering that the acceleration electric field $E_y$ is
on the same order as $B_0$
and considering that reconnection continues
until two outflow jets toward the $\pm x$-directions
meet each other at the periodic boundary at $x=\pm L_x/2$.

\section{Discussions}

\subsection{Parameter dependence and comparison}

\begin{figure}[htbp]
\begin{center}
\includegraphics[width={\columnwidth},clip]{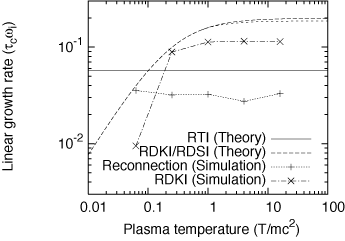}
\caption{Temperature dependency of the growth rates
($\tau_c\omega_i$).
Theoretical growth rates of RTI and RDKI, and
observed growth rates of magnetic reconnection and the RDKI
are presented.
\label{fig:rec_vs_dki}}
\end{center}
\end{figure}

Figure \ref{fig:rec_vs_dki} shows theoretical and observed
growth rates of simulation runs R1-R5 and D1-D5
with different temperature parameters $T/mc^2=1/16,1/4,1,4,16$.
The growth rates ($\omega_i$) are normalized by the light transit time $\tau_c$.
The linear growth rates of magnetic reconnection
are calculated from the reconnected field energy
$\sum B_z^2 $ in the neutral plane.
The normalized growth rates are around $\tau_c\omega_i \sim 0.03$.
Theoretical growth rates of reconnection are represented by
the theory of the relativistic tearing mode \citep{zelenyi79}.
Its growth rates $\gamma_{RTI}$ in pair plasmas are:
\begin{equation}
\tau_c \gamma_{RTI}=
\left\{ \begin{array}{ll}
\frac{1}{\sqrt{\pi}} ~ k\lambda (1-k^2\lambda^2) \beta^{3/2} ~ (2T/\gamma_{\beta}mc^2)^{-1/2} & (T \ll mc^2) \\
\frac{2\sqrt{2}}{\pi} ~ k\lambda (1-k^2\lambda^2) \beta^{3/2}& (T \gg mc^2) \\
\end{array} \right.
\label{eq:rti}
\end{equation}
In the relativistic case, we obtain the maximum growth rate by setting $k\lambda=1/\sqrt{3}$ 
\begin{equation}
\tau_c\gamma_{RTI} \lesssim 0.35 \beta^{3/2},
\label{eq:rti2}
\end{equation}
For $\beta=0.3$, the maximum growth rate is $\tau_c\gamma_{RTI} \sim 0.055$.
This rate is $1.7-2.0$ times faster than simulation results,
but remember that reconnection is not identical to the tearing mode. 
The linear growth rates of RDKI/DKI are obtained
from the perturbed magnetic field $\delta B_x$ in the neutral plane,
and the growth rates of the most dominant mode are selected.
The growth rates of the RDKI/RDSI are estimated by
equations \eqref{eq:dki4} and \eqref{eq:rdki5}.
The growth rate of the DKI increases as $T$ increases
in the nonrelativistic regime of $T/mc^2 \ll 1$.
On the contrary, the RDKI is insensitive to $T$.
The theory (eq. [\ref{eq:dki4}]) clearly explains this signature.
Relativistic pressure enhances effective inertia
through the fluid enthalpy term,
and then the enhanced inertia cancels further growth of the RDKI. 
So, the rate stays constant in the relativistic limit. 
Since equation \eqref{eq:dki5} slightly exceeds unity
($k_y\lambda \sim 1.1-1.3$)
in relativistic cases of $T \gtrsim mc^2$,
another criteria of equation \eqref{eq:rdki5} is presented
as a second branch of the theoretical rate
in Figure \ref{fig:rec_vs_dki}.
Both criteria give $\tau_c\gamma_{RDKI} \sim 0.19$,
which limits the obtained value of $\tau_c\omega_{i} \sim 0.11$.

\begin{figure}
\begin{center}
\includegraphics[width=0.7\columnwidth,clip]{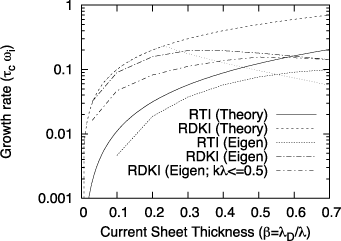}
\caption{\label{fig:rec_vs_dki_beta}
Growth rates of the instabilities as a function of the current sheet thickness
$\beta=\lambda_D/\lambda$.
The growth rate $\omega_i$ is presented in a timescale of $\tau_c$.
Upper limit of the RDKI $\beta$ (from eq. [\ref{eq:rdki7}]),
upper limit of the relativistic tearing mode $0.35 \beta^{3/2}$ (from eq. [\ref{eq:rti}]),
obtained eigen growth rates for the tearing mode,
obtained eigen growth rates for the RDKI
and
eigen growth rates for the RDKI for short wavelength are shown.
}
\end{center}
\end{figure}

Figure \ref{fig:rec_vs_dki_beta} shows
eigen and theoretical growth rates
as functions of the current sheet thickness
$\beta=\lambda_D/\lambda$ (eq. [\ref{eq:debye}]).
In this case, the plasma temperature is fixed to $T=mc^2$.
The upper limit of the RDKI $\beta$ (eq. [\ref{eq:rdki7}]),
the upper limit of the relativistic tearing mode
$0.35 \beta^{3/2}$ (eq. [\ref{eq:rti}]),
eigen growth rates for the relativistic tearing mode,
and eigen growth rates for the RDKI are presented.
For eigen growth rates for the RDKI,
we present two cases with $k\lambda=0.5$ and $k\lambda=1.0$
because we do not know what wavelength is the most dominant
and because the DKI theories use
the long-wavelength assumption ($k\lambda \lesssim 1$).
In the range of $0.03 < k\lambda < 0.7$,
it seems that
the eigen growth rates of the RTI and the RDKI/RDSI
are well limited by their theoretical upper limit.
It is still unclear
how the instabilities grow in the thick current sheet
and whether or not they are limited by the upper limits. 
This should be investigated by larger PIC simulations
or relativistic two-fluid MHD simulations.
Theoretically, equation \eqref{eq:upper} makes sense
as an upper limit of the RDKI,
because $\beta$ also represents the current (eq. [\ref{eq:curr}])
and because the RDKI is a current-driven instability.

Figure \ref{fig:rec_vs_dki2} compares
the theoretical growth rates
as functions of two parameters: $T/mc^2$ and $\beta$.
We use equations \eqref{eq:dki4}, \eqref{eq:rdki5}, and \eqref{eq:rti}. 
Generally, the RDKI/RDSI grows faster than reconnection.
Most parts of the frontmost flat region ($\beta \ll 1$ and $T\gg mc^2$)
rely on equation \eqref{eq:upper} or \eqref{eq:rdki5}.
It seems that the RDKI's rate goes down around $\beta \sim 1$,
but our fluid theory is no longer valid here,
because the current sheet is too thin (eq. [\ref{eq:debye}]).
Also, the fluid theory is not valid
around the backward region of $\beta > 0.1$ and $T/mc^2 \ll 1$,
where the Larmor radius is comparable with or larger than the sheet thickness $\lambda$ (eq. [\ref{eq:gyro}]).
Except for these invalid regions, 
we can conclude that the RDKI is more likely to occur
in a relativistic hot condition of $T/mc^2 \gtrsim 1$.

Importantly, in the relativistic regime of $T \gtrsim mc^2$
our simulations show no drastic change. 
The various aspects are similar:
the evolution of the reconnection or the RDKI structure,
time history of total energy, nonthermal ratio,
relative height of the nonthermal slope,
and the maximum energy of the nonthermal tail.
These results are quite reasonable considering the following facts.
The characteristic speed is around the light speed $\sim c$
(Table \ref{tab:rec_alfven}) and
the typical Larmor radius relative to the sheet width
is nearly constant (eq. [\ref{eq:rgyro}]).
It is straightforward that
the typical gyro period $(\Omega_c^{-1}=\gamma/\omega_c)$
is constant
when normalized by the light transit time $\tau_c=\lambda/c$.
So, we can conclude that the physical processes are ``similar''
in these relativistic cases, as long as we consider kinetics. 
Therefore, by employing
the moderately-relativistic simulation ($T \approx mc^2$),
we may figure out ultra-relativistic kinetic evolution,
which is computationally more expensive;
one has to resolve the gyro motion of the lowest energy particle ($\omega_{c}^{-1}$)
while the system evolution is governed by the typical gyro period
$(\Omega_c^{-1} = \gamma/\omega_{c})$. 
If we start to consider radiation effects,
we need another approach
because the synchrotron radiation highly depends on
the ultra relativistic Lorentz factor: $\gamma$ or $T/mc^2$.

\subsection{Pulsar wind problem}

Let us discuss a possible application to
the striped pulsar wind \citep{lyu01,kirk03}.
Our comparison shows that
the RDKI is the most likely process
in the current sheet,
as long as plasmas are relativistic hot ($T \gtrsim mc^2$).
If the current-aligned magnetic field
(the so-called guide field, $B_y$) exists,
secondary magnetic reconnection will dominate \citep{zeni05b};
however, magnetic fields are expected to be highly toroidal
inside the striped pulsar wind
due to the fast rotation of the neutron star,
and therefore the guide field would be weak or unlikely to exist.
Therefore, here we assume that
the RDKI dominates in the striped current sheets. 

\citet{kirk03} (hereafter KS03) has examined
the magnetic dissipation problem
by using the expanding current sheets model.
They estimated the magnetic dissipation rate,
assuming that striped current sheets are expanding with
a ``dissipation speed'' of $c\beta_c$,
where $\beta_c$ is the dimensionless speed. 
They assumed magnetic reconnection as the most likely process, and
they estimated the dissipation speed with the growth rate of
the relativistic tearing mode \citep{zelenyi79}
in their ``tearing mode-limited'' scenario.
Importantly, one-dimensional simulation results are
well described by an asymptotic analytic solution in all cases.
From the flux-freezing equation (eq. [38] in KS03),
the entropy equation (eq. [39] in KS03) and the dissipation speed ($\beta_c$),
the asymptotic solution $\Delta \propto (r/r_L)^{-q}$ can be obtained,
where $\Delta$ is a ratio of the current sheet thickness to
the wavelength of the striped current sheets
and $q$ is the spectral index. 
Two of their scenarios are presented in Table \ref{KS03table}.
They assume that the current sheets are
completely dissipated at the radial distance of $r_{max}/r_{LC}$
from the central neutron star,
where $r_{LC}$ is the distance of the light cylinder.
Two columns of $r_{max}/r_{LC}$ are presented;
the first column is the analytical form, and
the second column is the estimated value by using Crab parameters.
The termination shock is at $r/r_{LC}=2.0\times 10^9$, and so 
$r_{max}/r_{LC}<2.0\times 10^9$ is a favorable result.
The ``fast'' scenario is a physical upper limit;
its dissipation speed is based on the relativistic sound speed of $c/\sqrt{3}$.
In the tearing mode-limited scenario, 
they set $c\beta_c = \lambda \gamma_{RTI}$,
where $\tau_c\gamma_{RTI}=\beta^{3/2}$ (eq. [35] in KS03)
is a simplified form of equation \eqref{eq:rti}. 
If we modify the dissipation speed with
the maximum growth rate of $\tau_c\gamma_{RTI} \sim 0.35 \beta^{3/2}$ (from eq. [\ref{eq:rti2}]),
the dissipation distance ($r_{max}$) increases by 50\%.
The modified value is presented as
the ``tearing mode--limited (modified)'' scenario in Table \ref{KS03table}.

\begin{figure}[htbp]
\begin{center}
\includegraphics[width={0.7\columnwidth},clip]{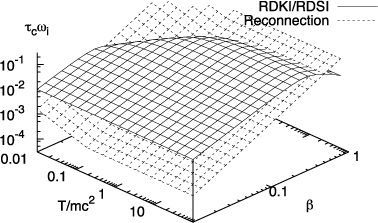}
\caption{Theoretical estimate of the growth rates $\omega_i$,
as a function of the plasma temperature $T/mc^2$ and the drift speed $\beta$.
The dashed surface shows linear growth rates of magnetic reconnection
in electron-positron plasmas, based on \citet{zelenyi79}.
The solid surface shows the upper limit
growth rates of the drift kink/sausage instability.
\label{fig:rec_vs_dki2}}
\end{center}
\end{figure}

We propose a similar dissipation scenario by the RDKI.
We consider that the current sheet evolves to
the broadened current sheet by the RDKI
and then it continues to expand by the cascading RDKI processes.
We call it the ``drift kink mode--limited'' scenario. 
We assume that the current sheet expands by $h \sim$(3-4) times
by single RDKI process:
the sheet's typical width $\lambda$ becomes
$\lambda' \sim h\lambda$,
where the prime denotes the physical value
in the broadened current sheet
after the first instability is saturated.
Consequently, the plasma density in the current sheet is
$n^{'}_{0} \sim n_0/h$.
For simplicity, we ignore the particle acceleration effect,
because nonthermal particles carry
less than 10\% of kinetic energies (in Fig. \ref{fig:dki_hist}\textit{b}).
Assuming the pressure balance condition
$2n_{0}T=2n'_{0}T'=B_{0}^2/8\pi$,
the second-stage plasma temperature increases to $T' \sim hT$
due to the nonlinear mixing of the RDKI.
Then, the Debye length
$(\lambda'_{D}/\lambda') \sim (\lambda_D / \lambda)$ and
the  Larmor radius $(r'_L/\lambda') \sim (r_L/\lambda)$
remain constant, relative to the sheet width.
Regarding the the relativistic Harris parameters,
$\beta' \sim \beta$ remains constant,
while the temperature increases, $T' \sim hT$.
According to Figure \ref{fig:rec_vs_dki2},
the RDKI is likely to occur again in the broadened condition.
So, it is quite plausible that
the current sheet is dissipated by series of the RDKI processes.
Topologically, the RDKI is favorable for the current sheet dissipation scenario,
because it does evolve into the broadened current sheet structure,
while it is not sure that reconnection evolves into the broaden current sheet. 
In thick current sheets of $\beta \ll 1$, we evaluate the dissipation speed
using equation \eqref{eq:rdki7}; $c\beta_c \sim \lambda \gamma_{RDKI} = c\beta$.
It is a good approximation of the eigen growth rates by a factor of 1-3
(Fig. \ref{fig:rec_vs_dki_beta}).
Using this, following KS03's method,
one can obtain the asymptotic index $q=2/5$ and relevant values.
These values are presented in Table \ref{KS03table}.
If we use the typical parameters,
the obtained dissipation distance is $r_{max} / r_L =3.1 \times 10^9$.
This value is 3 times better than
the original tearing mode scenario and
the drift-kink-mode scenario substantially improves
the dissipation model in the striped pulsar wind.

By the way, our DKI/RDKI theory is based on
a long-wavelength assumption.
In a nonrelativistic ion-electron plasma,
it is reported that short-wavelength current sheet instability
grows faster than long-wavelength modes \citep{suzuki02}.
Magnetic dissipation becomes even faster,
if short-wavelength mode grows faster in a thick current sheet.
In a thin current sheet,
the nonthermal accelerations are more likely to
transfer magnetic energy into the $y$-momentum $\pm p_{y}$
of nonthermal plasmas.
Then, since the $z$-pressure relatively decreases,
the current sheet may be thinner and
the RDKI may grow faster (Fig. \ref{fig:rec_vs_dki2}).
In a thick current sheet of $\beta \ll 1$,
we can probably ignore nonthermal acceleration
because few particles satisfy the threshold condition
(eq. [\ref{eq:threshold}]).
When neighboring current sheets come close to each other,
they start to collide each other
during the nonlinear stage of the RDKI.
We compare several runs
with different simulation box size $L_z$
(runs D3, D3b, and D3c in Table \ref{table}).
In run D3c, the current sheets collide with each other
across the periodic $z$-boundaries and
more than 90\% of the magnetic energy is
dissipated into particle energy. 
On the other hand, the actual dissipation speed
may be somewhat slower than
the growth rates of the instabilities
(both the RDKI and the tearing mode).
So, an effective dissipation speed
should be further investigated by a larger simulation,
in which we can observe cascading evolution of the RDKI.
If the current sheet structure survives at the termination shock,
the magnetic fields may further be dissipated by
the collision of the shock and current sheets.
In the downstream of the shock,
magnetic reconnection may be triggered
by turbulent electric fields.

In summary, we propose the RDKI
as an alternative magnetic dissipation scenario
in the striped current sheets,
based on our theoretical estimates.
At present we do not know whether or not
the RDKI is the final key to explain the $\sigma$ problem;
however, the drift kink mode scenario is obviously
more favorable than the reconnection/tearing mode scenario.

\subsection{Miscellaneous discussions}

In ion-electron reconnection,
various instabilities occur in different scales.
For example, the LHDI occurs on an electron scale,
the DKI on an ion scale,
reconnection is on a macroscopic MHD scale,
and then
coupling across these scales leads a complexity of plasma phenomena. 
In this study, we investigate a very thin current sheet
in $e^{\pm}$ plasmas.
Consequently, both the reconnection and the RDKI occur
in the same scale as the current sheet.
Both of them greatly violate the current sheet structure,
and we discuss them as competing processes.
When we consider a ``thick'' current sheet,
the RDKI or other cross-field instabilities may occur
on a rather microscopic scale,
and then they may interact with the macroscopic reconnection process.
These situations should be further investigated. 

It has been believed that
the reconnection's global energy dissipation is
controlled by the Hall physics \citep{birn01}.
In pair plasmas, the relevant positron's contribution
is canceled out by electron's, and so the Hall effect does not exist.
However, simulations of
pair plasma reconnection \citep{zeni01,claus04,zeni05b,bessho05}
reported fast reconnection rates
[$v_{in}/v_{out}$ or $E_y/(cB_0)$, $E_y/(v_AB_0)$] of $0.1$ or above.
This issue is equivalent to the question
``What is the origin of the reconnection electric field?''.
In classical nonrelativistic studies,
the electric field is described by
\begin{equation}
\vec{E}=-\vec{v}_e \times \vec{B} - \frac{1}{n_ee} \div \Pi_e
-\frac{m_e}{e}\Big(\frac{\partial \vec{v}_e}{\partial t} + \vec{v}_e \cdot \grad \vec{\vec{v}_e} \Big),
\label{eq:hesse94}
\end{equation}
where $v_e$ is the electron's velocity and
$\Pi_e$ is the electron's pressure tensor.
The Hall effect comes from the difference in
the electron Lorentz term ($\vec{v}_e \times \vec{B}$)
and its ion counterpart,
and it smears out in the case of pair plasmas.
\citet{bessho05} showed that
the pressure-tensor term explains the electric field
in the vicinity of the neutral region.
Similar analysis can be applied to relativistic cases,
using the relativistic pressure tensor \citep{write75}.
In addition, the electron's inertia may be of importance
due to the increasing mass from the Lorentz effect.

\subsection{Summary and Conclusion}

Let us summarize and conclude this paper.
We have carried out series of PIC simulations
of relativistic current sheet problems in pair plasmas.
We examine the relativistic magnetic reconnection
and we report that particles can be accelerated
near the magnetic pile-up region
as well as the $X$-type region.
We have also studied the current sheet instabilities,
which exists in the perpendicular plane to reconnection.
The parameter survey shows that
properties of these two processes are similar
in the relativistic regime of $T\gtrsim mc^2$,
as long as we consider the kinetics.
In addition, by comparing their growth rates,
the RDKI is the more likely process
in the relativistic regime of $T\gtrsim mc^2$.
Therefore we propose that
the magnetic dissipation in the striped pulsar wind
should be re-considered by using the RDKI.

\begin{acknowledgments}
The authors are grateful to
T. Terasawa, S. Abe, I. Shinohara, T. Yokoyama, S. Shibata,
K. Nagata, C. Jaroschek, and M. Hesse for fruitful discussions.
This work was supported by the facilitates of JAXA
and the Solar-Terrestrial Environment Laboratory, Nagoya University.
One of the authors (S. Z.) was granted a fellowship
from the Japan Society of the Promotion of Science and
GEST program of University of Maryland Baltimore County.
S. Z. is grateful to the editor and anonymous referee
for improving the original manuscript.
\end{acknowledgments}

\appendix

\newpage
\section{Nonthermal ratio parameter}

\begin{figure}[htbp]
\begin{center}
\includegraphics[width={0.5\columnwidth},clip]{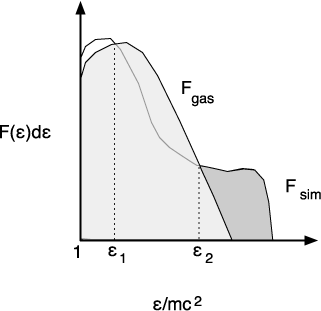}
\caption{Relation between the observed distribution $F_{sim}$ (\textit{white and dark gray}) and the equivalent thermal distribution $F_{gas}$ (\textit{light gray}). Two parameters $\varepsilon_1$ and $\varepsilon_2$ represent crossing points of the two spectra.
The high-energy tail region in dark gray
carries the ``nonthermal'' energy.
\label{fig:nonth_spec}}
\end{center}
\end{figure}

In this appendix, we describe our ``nonthermal ratio'' index,
to discuss the nonthermal acceleration.
By integrating over half of the simulation box,
we obtain the plasma energy spectra $F_{sim}(\varepsilon)$.
The total particle number $N$ and
the total kinetic energy $K$ satisfy
\begin{eqnarray}
N &=& \int_0^{\infty} d\varepsilon ~ F_{sim}(\varepsilon) \\
K &=& \int_0^{\infty} d\varepsilon ~ F_{sim}(\varepsilon) (\varepsilon - mc^2).
\end{eqnarray}
Next, we consider an ideal thermal gas,
which is described by the \JS distribution with unknown temperature $T$.
Similarly, we consider the particle number $N_{gas}$, the total kinetic energy $K_{gas}$
and its energy spectra $F_{gas}(\varepsilon)$ for this ideal gas.
The \JS gas has the relation \citep{synge}
\begin{equation}
K_{gas}/(N_{gas}mc^2)=3 (T/mc^2) + [K_1(mc^2/T)/K_2(mc^2/T)-1],
\label{eq:js}
\end{equation}
where $K_{1,2}(x)$ is the modified Bessel function of the second kind.
Then, we assume that $N_{gas}=N$ and $K_{gas}=K$,
so that $F_{gas}$ represents an ``equivalent'' thermal distribution.
Since equation \eqref{eq:js} is
a monotonically increasing function of $(T/mc^2)$,
we can find a unique solution of $T=T'$.
Then, we calculate the nonthermal parameter
by comparing the original spectrum $F_{sim}(\varepsilon)$
and the equivalent thermal spectrum $F_{gas}(\varepsilon)$ with $T=T'$.
The relation between the two spectra is illustrated in Figure \ref{fig:nonth_spec};
the spectrum $F_{gas}$ is presented in light gray, and
the spectrum $F_{sim}$ is presented behind it
in white and dark gray.
The two spectra cross at several points, e.g. 
$\varepsilon=\varepsilon_1,\varepsilon_2,...$, and
we define the nonthermal energy $K_{nth}$ as
\begin{equation}
K_{nth} = \int_{\varepsilon_2}^\infty [ F_{sim}(\varepsilon) - F_{gas}(\varepsilon) ] (\varepsilon-mc^2) ~d\varepsilon,
\end{equation}
where $F_{sim}$ exceeds $F_{gas}$ at
$\varepsilon=\varepsilon_2$ in the highest energy tail.
This ``nonthermal'' part is presented
in dark gray in Figure \ref{fig:nonth_spec}.
We use the ratio of the kinetic energy carried by this tail ($K_{nth}$) to
the total kinetic energy ($K$)
as the nonthermal ratio parameter.

\section{Linear analysis of relativistic drift kink/tearing mode}

We start with the relativistic fluid equation of motion,
\begin{equation}\label{eq:T}
T^{\alpha\beta}_{(fluid)} = ( \varepsilon + p ) u^{\alpha} u^{\beta} + p \eta^{\alpha \beta}
\end{equation}
where $T_{(fluid)}$ is the energy-momentum tensor for fluid,
$\varepsilon$ is the energy, $p$ is the pressure,
$u$ is the mean four-velocity of the fluid, and
$\eta^{\alpha \beta}$ is the metric tensor.
In order to preserve the energy and the momentum in the system,
\begin{equation}\label{eq:T1}
\frac{\partial}{\partial x^{\beta}} ( T^{\alpha\beta}_{(fluid)} + T^{\alpha\beta}_{(EM)} ) = 0
\end{equation}
should be satisfied, where $T_{(EM)}$
is the electromagnetic stress-energy tensor.
Using the relation
\begin{equation}\label{eq:T2}
\frac{\partial}{\partial x^{\beta}} T^{\alpha\beta}_{(EM)} = - F^{\alpha\beta} j_{\beta}
~,
\end{equation}
where $F$ is the electro-magnetic tensor and $j$ is the four current,
equation \ref{eq:T1} can be modified to
\begin{equation}\label{eq:T3}
\frac{\partial}{\partial x^{\beta}} T^{\alpha\beta}_{(fluid)} = F^{\alpha\beta} j_{\beta}
~.
\end{equation}
The spatial three components of equation \eqref{eq:T3}
can be described (See \cite{sakai}) as
\begin{equation}\label{eq:fluid}
\frac{\gamma^2_s}{c^2}
(p_s + \varepsilon_s) 
( \pp{}{t} + \vec{v}_s \cdot \grad )
\vec{v_s}
=
- \grad p_s +
\gamma_s q_s n_s 
\Big( \vec{E} +  \frac{\vec{v}_s}{c} \times \vec{B}  \Big)
- \frac{\vec{v}_s}{c^2} ( \gamma_s q_s n_s \vec{E} \cdot \vec{v}_s + \pp{p_s}{t} ),
\end{equation}
where the subscript $s$ denotes the species. $p$ for positrons and $e$ for electrons.
We also use the particle conservation law,
\begin{equation}\label{eq:cont}
\pp{}{t}(\gamma_s n_s) + \div(\gamma_s n_s \vec{v}_s ) = 0,
\end{equation}
and Maxwell equations,
\begin{equation}\label{eq:rotb}
\rot \vec{B} = \frac{ 4\pi }{c} \sum_s \gamma_s q_s n_s \vec{v}_s + \frac{1}{c} \pp{ \vec{E}}{t}
\end{equation}
\begin{equation}\label{eq:rote}
\rot \vec{E} =  - \frac{1}{c} \pp{ \vec{B}}{t}
\end{equation}
We assume the plasma is an adiabatic gas,
\begin{equation}\label{eq:ad}
p_s \propto n_s^{\Gamma}.
\end{equation}
where $\Gamma$ is the polytropic index of $\Gamma = 4/3$. 
Then, we replace some variables in equations
\eqref{eq:fluid}, \eqref{eq:cont}, \eqref{eq:rotb} and \eqref{eq:rote}
by using
\begin{equation}
\varepsilon_s = n_s m_0 c^2 + \frac{1}{\Gamma -1} p_s
\end{equation}
\begin{equation}
\gamma_s = [ 1-(\vec{v}_s/c)^{2} ]^{-1/2}
\end{equation}
\begin{equation}
d_s = \gamma_s n_s
\end{equation}
where $d_s$ stands for the number density in the observed frame.

Finally, equations
\eqref{eq:fluid}, \eqref{eq:cont}, \eqref{eq:rotb}, \eqref{eq:rote} and \eqref{eq:ad}
are linearized for the the Harris field model,
assuming that perturbations are given by
$\delta f \propto \delta f(z) \exp(ik_x x + ik_y y + \omega_i t)$.
The operators are transformed as
$\partial/\partial x \rightarrow i k_x $
and
$\partial/\partial y \rightarrow i k_y $,
and we normalize the first-order perturbations as
$\vec{B} = {B}_{0} ( F(z) \hat{x} + b_y \hat{y} ) + B_0 \hat{\vec{b}}$,
$\vec{E} = B_0 \hat{\vec{e}}$,
$\vec{v}_s = \pm c\beta + c\beta \hat{\vec{v}}_s$,
$d_s = d_{s0} f( z ) + d_{s0} \hat{d}_s$,
and
$p_s = p_{s0} f( z ) + p_{s0} \hat{p}_s$,
where
$F(z)=\thh$ and $f(z)=\cosh^{-2}(z/\lambda)$.
The parameter $b_y=B_y/B_0$, the amplitude of the guide field, is set to $0$. 
For numerical simplicity, some physical properties
are calculated in the form of summaries or differences between
the positron values and electron values,
$\hat{D}_{\pm} = \hat{d}_p \pm \hat{d}_e$,
$\hat{P}_{\pm} = \hat{p}_p \pm \hat{p}_e$,
and
$\hat{\vec{V}}_{\pm} = \hat{\vec{v}}_p \pm \hat{\vec{v}}_e$.
We have

\begin{eqnarray}
\pp{}{t} \hat{b}_x &=& - c k_y ( i\hat{e}_z ) + c \pp{}{z} \hat{e}_y \\
\pp{}{t} \hat{b}_y &=& - c \pp{}{z} \hat{e}_x + c k_x ( i\hat{e}_z ) \\
\pp{}{t} ( i \hat{b}_z ) &=& c k_x \hat{e}_y - c k_y \hat{e}_x \\
\pp{}{t}  \hat{e}_x &=& -A_1 f(z) \hat{V}_{x-}
+ c k_y (i \hat{b}_z) - c \pp{}{z} \hat{b}_y \\
\pp{}{t} \hat{e}_y &=& -A_1 ( \hat{D}_{+} + f(z) \hat{V}_{y-} )
+ c \pp{}{z} \hat{b}_x - c k_x ( i \hat{b}_z ) \\
\pp{}{t} ( i \hat{e}_z ) &=& - A_1 f(z) ( i\hat{V}_{z-} ) 
+ c k_y \hat{b}_x - c k_x \hat{b}_y \\
\pp{}{t} \hat{D}_+ &=& - c\beta k_y ( i \hat{D}_- )
- c\beta f(z) \Big( k_x ( i \hat{V}_{x+} ) + k_y ( i \hat{V}_{y+} ) + \pp{}{z} \hat{V}_{z+} \Big)
\nonumber \\
&&- c\beta f'(z) \hat{V}_{z+} \label{matrix:d1} \\
\pp{}{t} (i\hat{D}_-) &=& c\beta k_y \hat{D}_+
+ c\beta f(z) \Big( k_x \hat{V}_{x-} + k_y \hat{V}_{y-} - \pp{}{z} (i \hat{V}_{z-}) \Big)
\nonumber \\
&&- c\beta f'(z) (i\hat{V}_{z-}) \label{matrix:d2} \\
\pp{}{t} \hat{P}_+
&=&  \Gamma \pp{}{t} \hat{D}_{+}
- \Gamma \frac{\beta^2}{1-\beta^2}  \pp{}{t} \hat{V}_{y-}
\label{matrix:p1} \\
\pp{}{t} ( i \hat{P}_- )
&=&  \Gamma \pp{}{t} ( i \hat{D}_{-} )
- \Gamma \frac{\beta^2} {1-\beta^2} \pp{}{t}  (i \hat{V}_{y+} )
\label{matrix:p2} \\
\pp{}{t} ( i \hat{V}_{x+} ) &=&  c\beta k_y \hat{V}_{x-}
+ k_x \frac{A_2}{f(z)} \hat{P}_{+}
+ 2 A_3 ( i\hat{b}_{z} )
- A_3 b_y ( i\hat{V}_{z-} ) \\
\pp{}{t} \hat{V}_{x-} &=& - c\beta k_y ( i \hat{V}_{x+} )
- k_x \frac{A_2}{f(z)} ( i \hat{P}_{-} )
+ ( 2 A_3 / \beta ) \hat{e}_{x}
- A_3 b_y \hat{V}_{z+} \\
\pp{}{t} ( i \hat{V}_{y+} )
+ \frac{A_2 \beta}{c} \pp{}{t} ( i \hat{P}_{-} )
&=&
  c\beta k_y \hat{V}_{y-}
+ k_y \frac{A_2}{f(z)} \hat{P}_{+}
+ A_3 F(z) ( i\hat{V}_{z-} )
\label{matrix:v1} \\
\pp{}{t} \hat{V}_{y-} 
+ \frac{A_2 \beta}{c} \pp{}{t} \hat{P}_{+}
&=&
- c\beta k_y ( i \hat{V}_{y+} )
- k_y \frac{A_2}{f(z)} ( i \hat{P}_{-} ) 
+ 2 A_3 ( \beta^{-1} - \beta ) \hat{e}_{y}
\nonumber \\
&&+ A_3 F(z) \hat{V}_{z+}
\label{matrix:v2} \\
\pp{}{t} \hat{V}_{z+} &=& - c\beta k_y ( i \hat{V}_{z-} )
- \frac{A_2}{f(z)} \pp{}{z} \hat{P}_{+}
- \frac{A_3 F(z)}{f(z)} \hat{D}_{+} \nonumber \\
&&- 2 A_3 \hat{b}_{x}
+ A_3 b_y \hat{V}_{x-} - A_3 F(z) \hat{V}_{y-} \\
\pp{}{t} (i \hat{V}_{z-} ) &=& c\beta k_y \hat{V}_{z+}
-  \frac{A_2}{f(z)} \pp{}{z} ( i \hat{P}_{-} )
- \frac{A_3 F(z)}{f(z)} ( i\hat{D}_{-} )
\nonumber \\
&&+ ( 2 A_3 / \beta ) ( i \hat{e}_{z} )
+ A_3 b_y ( i\hat{V}_{x+} )  - A_3 F(z) ( i\hat{V}_{y+} )
\end{eqnarray}
where
$f'(z)= (\partial / \partial z) f(z) = -(2/\lambda) \thh \cs$,
$A_1 = 4 \pi e c\beta d_0 / B_0$,
$A_2 =({p_0}/{c\beta}) [ (\gamma_{\beta}^2/c^2) ( n_0 mc^2 + [\Gamma/(\Gamma-1)] p_0 ) ]^{-1}$,
and
$A_3 =({ed_0B_0}/c) [ (\gamma_{\beta}^2/c^2) ( n_0 mc^2 + [\Gamma/(\Gamma-1)] p_0 ) ]^{-1}$.
Note that equations
\eqref{matrix:p1}, \eqref{matrix:p2}, \eqref{matrix:v1} and \eqref{matrix:v2}
still contain time derivative terms.
We reorganize the six equations
\eqref{matrix:d1}, \eqref{matrix:d2},
\eqref{matrix:p1}, \eqref{matrix:p2},
\eqref{matrix:v1}, and \eqref{matrix:v2}
to obtain
$(\partial/\partial t) \hat{P}_+, (\partial/\partial t) ( i \hat{P}_- ),$
$(\partial/\partial t) ( i \hat{V}_{y+} )$, and
$(\partial/\partial t) \hat{V}_{y-}$.
\begin{eqnarray}
\left(\begin{array}{c}
\pp{}{t} \hat{P}_{+} \\
\pp{}{t} \hat{V}_{y-}
\end{array}\right)
= 
\frac{\Gamma}{1- \Gamma A_3 {\beta^3}{\gamma_{\beta}^2}/c}
\left(
\begin{array}{cc}
1 &- f(z) {\beta^2}\gamma_{\beta}^2 \\
-A_3 \beta /[cf(z)]& 1/\Gamma
\end{array}
\right)
\left(
\begin{array}{c}
rhs~of~\eqref{matrix:d1} \\
rhs~of~\eqref{matrix:v2}
\end{array}
\right)
\\
\left(\begin{array}{c}
\pp{}{t} \hat{P}_{-} \\
\pp{}{t} \hat{V}_{y+}
\end{array}\right)
= 
\frac{\Gamma}{1- \Gamma A_3 {\beta^3}{\gamma_{\beta}^2}/c}
\left(
\begin{array}{cc}
1 &- f(z) {\beta^2}\gamma_{\beta}^2 \\
-A_3 \beta /[cf(z)]& 1/\Gamma
\end{array}
\right)
\left(
\begin{array}{c}
rhs~of~\eqref{matrix:d2} \\
rhs~of~\eqref{matrix:v1}
\end{array}
\right)
\end{eqnarray}
Finally, after replacing $\partial/\partial t \rightarrow \omega_i$,
all the above relations can be solved
as an eigen value problem of the following matrix.
Solving this problem, we obtain the growth rate of the instability $\omega_i$ and
the $z$-structure of the perturbations for given $k_x$ and $k_y$.
The matrix is

\begin{equation}
\omega_i
\left(\begin{array}{c}
\hat{b}_x \\
\hat{b}_y \\
i\hat{b}_z \\
\hat{e}_x \\
\hat{e}_y \\
i\hat{e}_z \\
\hat{D}_+ \\
i\hat{D}_- \\
\hat{P}_+ \\
i\hat{P}_- \\
i\hat{V}_{x+} \\
\hat{V}_{x-} \\
i\hat{V}_{y+} \\
\hat{V}_{y-} \\
\hat{V}_{z+} \\
i\hat{V}_{z-} \\
\end{array}\right)
= 
\left(\begin{array}{ccccc}
a_{11} & a_{12} & \dots & \dots & a_{1n}\\
a_{21} & a_{22} & & & a_{2n} \\
\\
\\
\\
\\
\\
\\
\vdots &&\ddots & & \vdots \\
\\
\\
\\
\\
\\
\\
a_{n1} & a_{n2} & \dots & \dots & a_{nn} \\
\end{array}\right)
\left(\begin{array}{c}
\hat{b}_x \\
\hat{b}_y \\
i\hat{b}_z \\
\hat{e}_x \\
\hat{e}_y \\
i\hat{e}_z \\
\hat{D}_+ \\
i\hat{D}_- \\
\hat{P}_+ \\
i\hat{P}_- \\
i\hat{V}_{x+} \\
\hat{V}_{x-} \\
i\hat{V}_{y+} \\
\hat{V}_{y-} \\
\hat{V}_{z+} \\
i\hat{V}_{z-} \\
\end{array}\right).
\end{equation}
We set 241 grids in the $z$-direction and 20 grids per $\lambda$;
therefore, we solve $-6\lambda \le z \le 6 \lambda$.
The size of the matrix is $3856^2$.
The conductive-wall conditions are assumed at the boundaries.
For selected cases, we use 361 grids in $z$, $5776^2$ grids per a matrix
and we checked the effect of the grid resolution and the boundary positions.
In the case of $\beta=0.3$, the grid number shows an error less than 1\%.
In the case of $\beta<0.1$, we have to choose
a high resolution configuration with $5776^2$ grids. 
For numerical stability,
we sometimes add a small amplitude of the (nonrelativistic) viscosity term
to the fluid equation.
The equivalent Reynolds number ($R_e$) is set to
$R_e = \rho_0 \lambda c\beta / \mu = 10^6-10^7$. 
The eigen modes for the tearing mode or the RDKI/RDSI mode
are usually obtained without this viscosity term.
We need a small amplitude of the viscosity term to solve
short-wavelength RDKI modes ($k_x=0,k_y\gtrsim1$)
and the oblique modes ($k_x\ne 0,k_y\ne 0$). 

\clearpage

\clearpage

\clearpage

\clearpage

\clearpage

\clearpage

\clearpage

\clearpage

\clearpage

\begin{table*}
\begin{center}
\begin{tabular}{l|ccccc}
Run &
$T/mc^2$ &
$v_{in\_max}/c$ &
$v_{out\_max}/c$ &
$V_{A}/c$ &
$V_{A}^{inflow}/c$
\\
\hline
R1 &
16 &
$0.6$&
$0.97$&
$0.57$&
$0.99$
\\
R2 &
4 &
$0.4$&
$0.9$&
$0.57$&
$0.99$
\\
R3 &
1 &
$0.39$&
$0.88$&
$0.53$&
$0.98$
\\
R4 &
1/4 &
$0.29$&
$0.63$&
$0.45$&
$0.94$
\\
R5 &
1/16 &
$0.19$&
$0.53$&
$0.30$&
$0.84$
\\
\end{tabular}
\caption{\label{tab:rec_alfven}
Maximum outflow velocity compared with two types of the \Alfven velocity:
(1) the typical \Alfven velocity and (2) the inflow \Alfven velocity.
}
\end{center}
\end{table*}

\clearpage

\clearpage

\clearpage

\clearpage

\clearpage

\clearpage

\clearpage

\clearpage

\clearpage

\clearpage

\clearpage

\clearpage

\clearpage

\clearpage

\begin{deluxetable}{lccccc}
\tablecaption{\label{KS03table} Simple estimates of KS03 scenarios}
\tabletypesize{\scriptsize}
\tablewidth{0pt}
\tablehead{
\colhead{Dissipation} &
\colhead{q} &
\colhead{$r_{max}/r_{LC}$} &
\colhead{$r_{max}/r_{LC}$} &
\colhead{$\Delta(r/r_{LC})^{-q}$}
}
\startdata
Tearing-mode (modified) &
5/12 &
$0.75 \cdot \mu^{4/5} \hat{L}^{3/10}$&
$9.3 \times 10^9$&
$[ 5^8 0.35^2 /(19^7 12 \pi^2) ]^{1/12} \mu^{-1/3} \hat{L}^{-1/8}$
\\
Tearing-mode (KS03) &
5/12 &
$0.5 \cdot \mu^{4/5} \hat{L}^{3/10}$&
$6.1 \times 10^9$&
$[ 5^8/(19^7 12 \pi^2) ]^{1/12} \mu^{-1/3} \hat{L}^{-1/8}$
\\
Drift-kink-mode &
2/5 &
$0.25 \cdot \sqrt{\pi} \mu \hat{L}^{1/4}$&
$3.1 \times 10^9$&
$4^{-3/5} \pi^{1/5} \mu^{-2/5} \hat{L}^{-1/10}$
\\
Fast (KS03) &
$1/3$ &
$0.1 \mu^{2} (1-\beta_c^2)/\beta_c$ &
$4.6 \times 10^7$&
$\{ 6 \beta_c / [ 25\pi (1-\beta_c^2)] \}^{1/3} \mu^{2/3}$
\enddata
\tablecomments{
The dissipation index $q$,
the dissipation distance $r_{max}$,
its estimated value,
and the factor for $\Delta$
are shown.
The Crab parameters
$\mu=2 \times 10^4$ and $\hat{L}=1.5 \times 10^{22}$ 
are the total energy carried by the wind per unit rest mass (eq. [23] in KS03) and
the dimensionless flow luminosity (eq. [29] in KS03),
respectively.
}
\end{deluxetable}

\clearpage

\begin{deluxetable}{l|cccccccccccc}
\tabletypesize{\scriptsize}
\rotate
\tablecaption{\label{table} List of simulation runs}
\tablewidth{0pt}
\tablehead{
\colhead{Run} &
\colhead{$T/mc^2$} &
\colhead{$L_x$} &
\colhead{$L_y$} &
\colhead{$L_z$} &
\colhead{$c/\omega_{c}$} &
\colhead{$c/\omega_p$} &
\colhead{$\frac{n_{bg}}{(\gamma_{\beta}n_{0})}$} &
\colhead{$n_{total}$} &
\colhead{${K_{nth}}/{K}$} &
\colhead{$\tau_c \omega_i$} &
\colhead{$\varepsilon_{max}/mc^2$} &
\colhead{$\varepsilon_{est}/mc^2$}
}
\startdata
R1 &
16 &
102.4 &
-- &
51.2 &
0.01 &
0.077 &
5\% &
$1.7 \times 10^{7}$ &
$>3.5 \times 10^{-1} $ &
$3.3 \times 10^{-2}$&
$>3.4 \times 10^{3}$ &
$5.6 \times 10^{3}$
\\
R2 &
4 &
102.4 &
-- &
51.2 &
0.04 &
0.15 &
5\% &
$1.7 \times 10^{7}$ &
$>3.7 \times 10^{-1} $ &
$2.8 \times 10^{-2}$ &
$>6.9 \times 10^{2}$ &
$1.4 \times 10^{3}$
\\
R3 &
1 &
102.4 &
-- &
51.2 &
0.16 &
0.31 &
5\% &
$1.7 \times 10^{7}$ &
$>3.7 \times 10^{-1} $ &
$3.3 \times 10^{-2}$ &
$>1.5 \times 10^{2}$ &
$3.5 \times 10^{2}$
\\
R4 &
1/4 &
102.4 &
-- &
51.2 &
0.63 &
0.61 &
5\% &
$1.7 \times 10^{7}$ &
$>3.4 \times 10^{-1} $ &
$3.2 \times 10^{-2}$ &
$>4.0 \times 10^{1}$ &
$8.7 \times 10^{1}$
\\
R5 &
1/16 &
102.4 &
-- &
51.2 &
2.5 &
1.2 &
5\% &
$1.7 \times 10^{7}$ &
$>1.1 \times 10^{-1}$ &
$3.6 \times 10^{-2}$ &
$>1.1 \times 10^{1}$ &
$2.2 \times 10^{1}$
\\
\hline
D1 &
16 &
-- &
25.6&
51.2 &
0.01 &
0.077 &
0 &
$4.2 \times 10^{6} $ &
$4.7 \times 10^{-2}$ &
$1.1 \times 10^{-1}$ &
$7.0 \times 10^{2}$ &
$7.1 \times 10^{2}$
\\
D2 &
4 &
-- &
25.6&
51.2 &
0.04 &
0.15 &
0 &
$4.2 \times 10^{6} $ &
$5.8 \times 10^{-2}$ &
$1.2 \times 10^{-1}$ &
$1.9 \times 10^{2}$ &
$1.6 \times 10^{2}$
\\
D3 &
1 &
-- &
25.6&
51.2 &
0.16 &
0.31 &
0 &
$4.2 \times 10^{6} $ &
$7.2 \times 10^{-2}$ &
$1.1 \times 10^{-1}$ &
$4.8 \times 10^{1}$ &
$4.2 \times 10^{1}$
\\
D4 &
1/4 &
-- &
25.6&
51.2 &
0.63 &
0.61 &
0 &
$4.2 \times 10^{6} $ &
$2.6 \times 10^{-2}$ &
$8.9 \times 10^{-2}$ &
$10.0$ &
$9.9$
\\
D5 &
1/16 &
-- &
25.6&
51.2 &
2.5 &
1.2 &
0 &
$4.2 \times 10^{6} $ &
N/A &
$9.5 \times 10^{-3}$ &
$4.0$ &
$3.0$
\\
\hline
R0 &
$\sim$ 1/4 &
102.4 &
-- &
51.2 &
0.54 &
0.54 &
1\% &
$6.7 \times 10^{7}$ &
$-$ &
$-$ &
$-$ &
$-$ \\
D3a &
1 &
-- &
25.6&
51.2 &
0.16 &
0.31 &
5\% &
$4.2 \times 10^{6} $ &
$3.0 \times 10^{-2}$ &
$1.1 \times 10^{-1}$ &
$4.5 \times 10^{1}$ &
$4.2 \times 10^{1}$
\\
D3b &
1 &
-- &
25.6&
38.4 &
0.16 &
0.31 &
0 &
$4.2 \times 10^{6} $ &
- &
- &
- &
-
\\
D3c &
1 &
-- &
25.6&
25.6&
0.16 &
0.31 &
0 &
$2.1 \times 10^{6} $ &
- &
- &
- &
-
\enddata
\tablecomments{
Initial parameters and obtained physical values of simulation runs in this paper.
Runs R0 and R1-R5 are for magnetic reconnection.
Runs D1-D5, D3a-D3c are for the DKI.
The temperature ($T/mc^2$), the system size
($L_x,  L_y, L_z$ in units of $\lambda$),
the unit gyroradius (in units of $\lambda$),
the unit inertial length (in units of $\lambda$),
the ratio of the background plasma density ($n_{bg}$)
to the plasma density in the current sheet ($\gamma_{\beta}n_{0}$),
the total number of the particles ($n_{total}$).
the nonthermal ratio ($K_{nth} / K$),
the linear growth rate ($\tau_c \omega_i$),
the maximum energy ($\varepsilon_{max} /mc^2)$,
and
their estimated values ($\varepsilon_{est} /mc^2)$
are listed.
}
\end{deluxetable}

\end{document}